\documentclass[default, iicol, sn-aps]{sn-jnl}

\usepackage{graphicx}%
\usepackage{multirow}%
\usepackage{amsmath,amssymb,amsfonts}%
\usepackage{amsthm}%
\usepackage{mathrsfs}%
\usepackage[title]{appendix}%
\usepackage{xcolor}%
\usepackage{textcomp}%
\usepackage{manyfoot}%
\usepackage{booktabs}%
\usepackage{algorithm}%
\usepackage{algorithmicx}%
\usepackage{algpseudocode}%
\usepackage{listings}%
\usepackage{dsfont}
\usepackage{verbatim}
\usepackage{bm}

\begin{document}

\title[Article Title]{Reservoir as a Dynamical System: A Tutorial Review}

\author[1]{\fnm{Sarika} \sur{Jalan}}
\author[2,3]{\fnm{Ling} \sur{Feng}}
\author*[1]{\fnm{Dishant} \sur{Sisodia}}\email{dishantsisodia.physics@gmail.com} 
\author[4]{\fnm{Kohei} \sur{Nakajima}}

\affil[1]{\orgdiv{Complex Systems Lab, Department of Physics}, \orgname{Indian Institute of Technology Indore}, \orgaddress{\street{Khandwa Road, Simrol}, \city{Indore}, \postcode{453552}, \country{India}}}
\affil[2]{\orgdiv{Institute of Advanced Intelligence and Computing (IAIC)}, \orgname{Agency for Science, Technology and Research (A*STAR)}, \postcode{138632},
\orgaddress{\country{Singapore}}}
\affil[3]{\orgdiv{Department of Physics}, \orgname{National University of Singapore}, \postcode{117551},
\orgaddress{\country{Singapore}}}
\affil[4]{\orgdiv{Graduate School of Information Science and Technology}, \orgname{The University of Tokyo}, \orgaddress{\street{7-3-1 Hongo, Bunkyo-ku}, \city{Tokyo}, \postcode{113-8656}, \country{Japan}}}

\abstract{This brief tutorial review examines the operation of reservoir computing  from the standpoint of nonlinear dynamics and provides a concise overview of recent advances in the field. We begin by delineating the predictive capabilities of reservoir systems, followed by a succinct yet sufficient introduction to the nonlinear dynamical systems techniques that will be employed throughout this article to elucidate reservoir functioning. Subsequently, we describe reservoir operation using these techniques in the context of successful prediction, accompanied by all necessary implementation codes. The objective of this article is to offer a practical guide to the implementation of reservoir computing while, from a nonlinear dynamics perspective, explaining the mechanisms by which reservoirs achieve accurate predictions.}

\maketitle

\tableofcontents
\section{Introduction} 
Over the recent decades, artificial intelligence and machine learning have progressed from niche computational techniques to a central paradigm for scientific discovery. 
In contemporary scientific computing, the field is increasingly shaped by the fusion of these data-driven approaches with classical mathematical models. As a result, machine learning is now firmly embedded in many branches of dynamical systems research \cite{10.1063/5.0016505, PhysRevLett.120.024102, Brunton2016, Bury2023, Choudhary2020, 10.1063/1.5120538, Tsironis2025, PhysRevE.105.054203}.
Reservoir computing (RC) is one such machine learning framework whose functioning can be interpreted through the mathematical framework of dynamical systems, making it a natural platform for investigating the interface of AI and the physics of nonlinear dynamics. RC is grounded in the architecture of recurrent neural networks (RNNs) \cite{LUKOSEVICIUS2009127,doi:10.1126/science.1091277} and has gained wide popularity, particularly because of its effectiveness for processing sequential data \cite{PhysRevResearch.3.013090,10.1063/1.4979665,PhysRevE.98.052209,nakajima2021reservoir}.
Addressing the demand for accessible pedagogy, recent efforts such as the RC bootcamp \cite{10.1063/5.0283386} employ Jupyter notebooks to guide learners from introductory Python programming to more advanced concepts in reservoir computing.
This review, however, sets out to move beyond the algorithmic viewpoint. Instead of considering RC solely as an efficient “black-box’’ machine learning tool, we approach it from the standpoint of dynamical systems theory, analyzing the reservoir as a high-dimensional driven dynamical system. Although multiple frameworks have been proposed to understand why RC works, such as information theoretic perspective \cite{Inubushi2017, Carroll2020, Mizzi2025}, mean field theory \cite{PhysRevE.87.042809}, universal approximation theorem \cite{GRIGORYEVA2018495}, and links to vector autoregression \cite{Bollt2021}, this tutorial article concentrates exclusively on the nonlinear dynamical systems perspective. While this article focuses exclusively on reservoir computing, we provide a brief overview of long short-term memory (LSTM) networks which is another prominent variation of RNNs, for the sake of completeness.

\section{RC: What it is ?}
Emerging from the broader RNN framework, RC is naturally well suited for handling sequential and temporal data \cite{doi:10.1126/science.1091277}. Conventional RNNs are hard to train with standard gradient descent, since the iterative reduction of training error is computationally intensive and leads to prohibitively long training times. To overcome this drawback, RC was proposed as a computational paradigm \cite{LUKOSEVICIUS2009127} that unifies independently developed models such as echo state networks and liquid state machines \cite{10.1016/j.neunet.2007.04.001}. Its key distinction from traditional RNNs lies in the training of intermediate (reservoir) layer; this will become clearer in the schematic description of RC operation that follows. At this stage, however, it is worth briefly highlighting RC’s advantage over earlier, more conventional architectures. The internal connections of the reservoir are kept entirely fixed, so that only the weights of the final readout layer are subject to optimization. By relocating the learning effort away from the recurrent core, RC sidesteps the typical training bottlenecks associated with gradient descent.

At its core, the RC machine maps a low $n$ dimensional input channel $u(t)$ into a higher $m$ dimensional space through a weight matrix $W_{in}$. The reservoir itself is represented by a network whose connectivity is encoded in the adjacency matrix $\mathcal{A}\in\mathbb{R}^{m\times m}$. This recurrent structure endows the reservoir with a memory, allowing its current state to retain information about past inputs. Both $W_{in}$ and $\mathcal{A}$ are initialized prior to training and remain fixed thereafter, which is a defining characteristic of RC. Typically, the elements of $W_{in}$ are drawn from a uniform distribution in the interval $[-b,b]$. A common choice for the reservoir topology is an Erd\H{o}s-R'enyi random network \cite{erdos1959random}, characterized by a connection probability $\sigma$. The resulting adjacency matrix is subsequently rescaled to achieve a prescribed spectral radius $\rho$, such that the magnitude of its largest eigenvalue equals $\rho$. This scaling plays an important role in controlling the dynamical behavior and memory capacity of the reservoir \cite{6706899}. The reservoir state $r[i]$ is updated by the following equation for $N_t$ time steps:
\begin{equation}
    r[i+1] = (1-\alpha)r[i] + \alpha \hspace{0.1cm} tanh(A r[i] + W_{in} u[i+1]), \nonumber
\end{equation}
where $\alpha$ is known as the leakage rate. The update rule combines information from the previous reservoir state with the current input, while $\alpha$ determines the relative contribution of these two components. Consequently, the reservoir dynamics reflect both the intrinsic memory of the network and the influence of the incoming signal.
During the training stage, the objective is to determine an output mapping that minimizes the discrepancy between the predicted and target outputs. To achieve this, the updated reservoir states are stored and stacked to form a matrix $\mathcal{R}$ of dimensions $(m \times N_t)$. Simultaneously, the target output data are assembled into a matrix $\mathcal{U}\in\mathbb{R}^{n\times N_t}$. The output weight matrix $W_{out}$ is then obtained through ridge (Tikhonov) regression:
\begin{equation}
    W_{out} = \mathcal{U}.\mathcal{R}^{T}.(\mathcal{R}\mathcal{R}^{T} + \beta \mathcal{I})^{-1}. \nonumber
\end{equation}
Once training is completed, the reservoir can operate autonomously. In this testing phase, the output generated at each time step is fed back as the input for the next step, and the predicted output is obtained as $v[i+1] = W_{out}r[i]$.

\begin{figure}[t!]
\includegraphics[width = \linewidth]{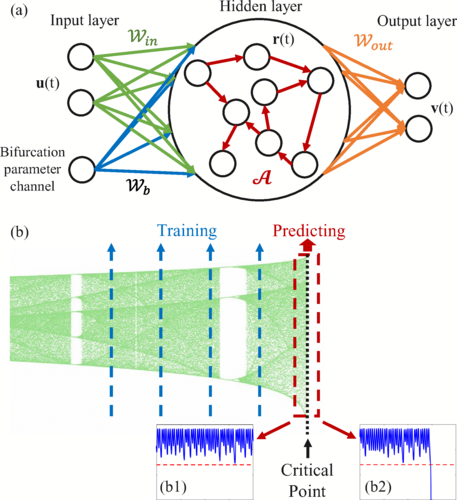}
\caption{Reused from \cite{PhysRevResearch.3.013090}(a) Schematic illustration of the parameter-aware reservoir computing architecture. In addition to the dynamical input $u$, an auxiliary parameter channel $\varepsilon$ is supplied to the reservoir. The matrices $W_{in}$ and $W_b$ project the dynamical input and parameter input, respectively, into the high-dimensional reservoir space. The recurrent interactions within the reservoir are governed by the adjacency matrix $\mathcal{A}$, while the trained output matrix $W_{out}$ generates the predicted output. During autonomous prediction, the output produced at the current time step is fed back as the input for the subsequent step. (b) Training with time series taken from a few parameter values in the green regime in which the system operates normally, indicated by the three vertical blue dashed lines. After training, prediction can be made either for parameter values before a critical point or after.}
\label{fig1}
\end{figure}
As illustrated in Fig.~\ref{fig1}, the  ``Parameter-aware architecture'' extends the conventional RC framework by incorporating an additional input channel that explicitly provides information about the system parameter. This modification enables the reservoir to learn not only the temporal evolution of the dynamics but also their dependence on the control parameter. Therefore, the dynamical evolution of the reservoir states \cite{PhysRevResearch.3.013090} is described as follows:
\begin{equation}\label{eq1}
\begin{aligned}
    r[i+1] =& (1-\alpha)r[i]  +  \alpha \hspace{0.1cm} tanh(A r[i] + \\&W_{in} u[i+1] + k_{b} W_b (\varepsilon - \varepsilon_b)).
\end{aligned}
\end{equation}
Here, $\varepsilon$ is the bifurcation parameter, $k_b$ and $\varepsilon_b$ are the hyperparameters. $W_b$, initially chosen from a uniform random distribution ranging from $[-c,c]$, remains fixed throughout the training. This gives us the following set of hyperparameters which can be tuned and optimized, $(m, b, c, \alpha, \rho, \sigma, k_b, \varepsilon_b)$. The reservoir is trained using data corresponding to multiple values of $\varepsilon$ prior to the critical transition. After training, the model is capable of accurately reproducing the system dynamics in the neighborhood of the transition, including parameter regimes both before and beyond the transition point.

\section{RC: Why is it popular ?}
RC has emerged as one of the most successful machine learning frameworks for modeling nonlinear dynamical systems. Since its introduction, 
the method has attracted considerable attention across diverse disciplines apart from nonlinear dynamics, including fluid dynamics \cite{PhysRevFluids.5.113506}, climate science \cite{https://doi.org/10.1029/2020MS002290}, neuroscience \cite{srinivasan2025boosting}, finance \cite{SANTOS2025134698}. The popularity of RC largely stems from its ability to learn complex temporal and nonlinear patterns while requiring the optimization of only the output layer, resulting in significantly lower training costs than conventional RNNs. Consequently, RC provides an attractive balance between predictive performance, computational efficiency, and implementation simplicity.

\paragraph{(i) Predictive power of RC for complex data:}
A large body of literature has demonstrated the remarkable predictive capabilities of RC in nonlinear and high-dimensional systems. Early studies established that RC can accurately reconstruct and forecast the dynamics of chaotic systems, when the underlying governing equations are unknown \cite{10.1063/1.4979665, 10.1063/1.5039508, PhysRevE.98.052209, shrimali, mandal}. The possibility of using virtual nodes as reservoir was successfully demonstrated using a logistic map \cite{PhysRevE.110.034204}. Beyond low-dimensional systems, RC has also been applied to high-dimensional spatiotemporal processes \cite{PhysRevLett.120.024102, li2025reservoir, alam2026attractor}, stochastic dynamics \cite{10.1063/5.0252908} often achieving predictive accuracies comparable to or exceeding those of more computationally demanding machine learning approaches.

\paragraph{(ii) Predictive power of PARC for chaotic and transitions regimes:}
While conventional RC achieved considerable success in forecasting nonlinear dynamical systems, predicting behavior near bifurcation points and critical transitions remained challenging. In these regimes, even small variations in system parameters can produce qualitative changes in the underlying dynamics, limiting the extrapolation capabilities of many data-driven approaches. To address this issue, parameter-aware reservoir computing (PARC) was introduced \cite{PhysRevResearch.3.013090}. It incorporates the control parameter directly into the reservoir as an additional input channel. This modification enables the framework to learn not only the temporal evolution of the system but also its dependence on the bifurcation parameter. Consequently, PARC has been shown to accurately identify critical transition points and reproduce the system dynamics in their vicinity, even when trained solely on data from pre-transition parameter regimes \cite{PhysRevE.104.014205,PhysRevResearch.3.013090, PhysRevResearch.6.043194, kong2023reservoir, panahi2024adaptable} along with adaptive control \cite{10.1063/5.0291585}, predicting dynamical states of a network \cite{10.1063/5.0271356} and model free prediction of multistability \cite{10.1063/5.0119963}. Such capabilities are particularly valuable in real-world settings, where the governing equations are often unavailable and slowly varying external influences may drive system parameters across bifurcation thresholds, potentially leading to undesirable phenomena such as amplitude death \cite{SAXENA2012205,KOSESKA2013173}, tipping \cite{ashwin2012tipping} or crises \cite{GREBOGI1983181}.

\paragraph{(iii) Brief comparison with previous algorithms:}
A variety of machine learning architectures have been applied to forecasting nonlinear dynamical systems, including feedforward neural networks (FFNNs), RNNs, long short-term memory (LSTM) networks, gated recurrent units (GRUs), and, more recently, transformer-based models. Although these approaches often achieve high predictive accuracy, they typically require large training datasets, substantial computational resources, and extensive hyperparameter tuning. In contrast, reservoir computing offers a significantly simpler training procedure by optimizing only the output layer while maintaining competitive predictive performance. Comparative studies have shown that RC can match or outperform RNNs, GRUs, LSTMs, transformers, and FFNN-based approaches in a range of time-series forecasting tasks \cite{SHAHI2022100300,pmlr-v199-cisneros22a,npg-27-373-2020}. Nevertheless, RC is not without limitations. Its predictive performance can be highly sensitive to the choice of reservoir hyperparameters, whose optimal selection is often nontrivial. Moreover, long-term forecasting of strongly chaotic systems remains challenging because of their inherent sensitivity to initial conditions.

\section{Methods and Techniques for dynamical systems analysis of RC}\label{sec:tools}
In this section we will equip the readers with various nonlinear dynamics techniques which will be used later for describing the functionality of RC.

\subsection{Eigenvalues of Jacobian:}\label{subsec:eig}
One of the most popular concepts in dynamical system is stability analysis; that is, how a dynamical system will react in presence of a ``small" perturbation. For this, we construct the Jacobian matrix of a given dynamical system whose eigenvalues provide a complete information of stable and unstable manifolds of the underlying system. 
Let us consider a general $d$-dimensional discrete-time dynamical system (map) defined by:
\begin{equation}
    \mathbf{x}_{t+1} = \mathbf{f}(\mathbf{x}_t , \mu), \nonumber
\end{equation}
where $\mathbf{x}_t \in \mathbb{R}^d$ is the state vector at time $t$, $\mu$ is a one-dimensional parameter and $\mathbf{f}: \mathbb{R}^d \rightarrow \mathbb{R}^d$ is a continuously differentiable vector valued function. Let $\mathbf{x}^*$ be a fixed point of the map, satisfying the condition $\mathbf{x}^* = \mathbf{f}(\mathbf{x}^* , \mu)$. To determine the local stability of this fixed point, we introduce a small perturbation $\mathbf{\delta x}_t$ around $\mathbf{x}^*$. Substituting the perturbed state expression back into the original map yields:
\begin{equation}
    \mathbf{x}^* + \mathbf{\delta x}_{t+1} = \mathbf{f}(\mathbf{x}^* + \mathbf{\delta x}_t). \nonumber
\end{equation}
Taylor series expansion around the fixed point $\mathbf{x}^*$, yields:
\begin{equation}
    \mathbf{f}(\mathbf{x}^* + \mathbf{\delta x}_t\;, \mu) = \mathbf{f}(\mathbf{x}^*, \mu) + D\mathbf{f}(\mathbf{x}^*) \mathbf{\delta x}_t + \mathcal{O}(\|\mathbf{\delta x}_t\|^2), \nonumber
\end{equation}
where $D\mathbf{f}(\mathbf{x}^*)$ represents the $d \times d$ Jacobian matrix evaluated at the fixed point, and $\mathcal{O}(\|\mathbf{\delta x}_t\|^2)$ denotes the higher-order nonlinear terms. Since $\mathbf{x}^* = \mathbf{f}(\mathbf{x}^*, \mu)$, the zero-order terms on both the sides cancel out. Neglecting the higher-order terms, we arrive at the linearized system $\mathbf{\delta x}_{t+1} = \mathbf{J} \mathbf{\delta x}_t$, where $\mathbf{J} = D\mathbf{f}(\mathbf{x}^*)$ is the Jacobian matrix, whose elements are defined explicitly by:
\begin{equation}
    \mathbf{J}_{ij} = \left. \frac{\partial f_i}{\partial x_j} \right|_{{\mathbf{x} = \mathbf{x}^*}}. \nonumber
\end{equation}
The magnitude of the eigenvalues of Jacobian dictates the phase space dynamics near the fixed point:
\begin{itemize}
    \item \textbf{Asymptotically Stable:} If $|\lambda_i| < 1$ for all $i$ (all eigenvalues lie strictly inside the unit circle in the complex plane).
    \item \textbf{Unstable:} If $|\lambda_i| > 1$ for at least one $i$ (at least one eigenvalue lies outside the unit circle).
    \item \textbf{Marginally Stable / Bifurcation Threshold:} If $|\lambda_i| = 1$ for some $i$ (eigenvalues on the unit circle).
\end{itemize}
For one-dimensional parameter variations in discrete-time dynamical maps, there are three generic ways in which eigenvalues can cross the unit circle, each corresponding to a distinct local bifurcation \cite{Kuznetsov:2008}. First, when a single real eigenvalue crosses the unit circle through $\lambda_i=1$ at $\mu=\mu_c$, the system undergoes a saddle-node bifurcation. Second, if a single real eigenvalue crosses through $\lambda_i=-1$ at $\mu=\mu_c$, a period-doubling bifurcation occurs. Finally, when a complex conjugate pair of eigenvalues crosses the unit circle, i.e., $\lambda_{i,j}=e^{\pm i\theta}$ at $\mu=\mu_c$, the system experiences a Neimark-Sacker bifurcation. 

\subsection{Asymptotic distance under perturbation}
\label{sec:asymp-distance}
As a complement to the Jacobian based diagnostics above, there exists
a global indicator, and numerical estimation of  this asymptotic distance does not require calculation of the derivatives. The idea is the oldest
one in the study of chaos: run two copies of a system from an ``almost
identical" set of initial conditions and watch whether they stay together or drift apart \cite{strogatz2015nonlinear,HIRSCH2013329,eckmann1985ergodic}, a property referred as sensitive dependence of initial conditions and is the hallmark of chaos. 
Treating a system as a discrete-time map,
\begin{equation}
  \bm{k}_{t+1} = \bm{F}(\bm{k}_{t}),
  \label{eq2}
\end{equation}
with state $\bm{k}_t$ and nonlinear operator $\bm{F}$, we seed a second
trajectory $\bm{k}_0' = \bm{k}_0 + \bm{\varepsilon}$ with a small (e.g.\
Gaussian) perturbation, iterate both for a large number of steps $T$, and
record the \emph{asymptotic distance}
\begin{equation}
  D = \bigl| \bm{k}_T' - \bm{k}_T \bigr|. \nonumber
  \label{eq:asymp-dist}
\end{equation}
In regular systems, perturbations decay with time, resulting in $D \to 0$, whereas in the chaotic state, nearby trajectories diverge. Consequently, $D$ provides a measure of the cumulative growth of an initial perturbation along the trajectory.
Since the estimate obtained from a single trajectory is susceptible to fluctuations, the asymptotic distance is averaged over an ensemble of $M$ input trajectories. As the distances typically span several orders of magnitude, the geometric mean provides a natural measure of the average,
\begin{equation}
  \widetilde{D} \;=\; \ln\!\Bigl[\,\bigl(D_1' D_2' \cdots D_M'\bigr)^{1/M}\Bigr]
              \;=\; \frac{1}{M}\sum_{i=1}^{M} \ln D_i' . \nonumber
  \label{eq:Dtilde}
\end{equation}
In the non chaotic regime, trajectories typically converge to within the floating point rounding error of the computer. Consequently, the minimum measurable value of $D$ is limited by the machine precision rather than being exactly zero. This also prevents the singularity $\ln D=\ln 0\rightarrow -\infty$, which would otherwise corrupt the average. A small baseline $D_i' = D_i + \delta$, is therefore added to $D$ before computing the logarithm with $\delta$ at the scale of the machine precision. The LSTM experiments ~\cite{wei2026multiple} later use $\delta = e^{-15}$, making $\ln D_i' = -15$ the effective ``numerical zero'' for the ordered phase.
The asymptotic distance should be interpreted with two important caveats. First, since the comparison is performed only at the final iteration, $\widetilde{D}$ serves as an indicator of the regular to chaotic transition rather than a quantitative measure of the divergence rate, which is instead characterized by the Lyapunov exponent. Second, in the chaotic regime, the separation between trajectories eventually saturates at a value determined by the size of the chaotic attractor rather than the instability strength. Consequently, a weakly chaotic system with a large attractor may exhibit a larger value of $D$ than a strongly chaotic system with a smaller attractor. Therefore, the asymptotic distance is most appropriately used as a complementary indicator of the order-chaos transition alongside Jacobian eigenvalue and bifurcation analyses.


\subsection{Crisis and bifurcation}\label{subsec:bif}
To establish a framework for defining a crisis, we first review essential concepts such as chaotic attractors and stable and unstable manifolds.
A map $f:I\rightarrow I$, where $I=[\alpha,\beta]$, is said to be chaotic  if it satisfies the following conditions \cite{HIRSCH2013329}:
\begin{enumerate}
    \item The periodic points of $f$ are dense in $I$. A set $S\subseteq I$ is said to be \emph{dense} in $I$ if, for every $x\in I$ and $\varepsilon>0$, there exists a point $y\in S$ such that
    $|x-y|<\varepsilon$. 
    
    \item $f$ is \textit{topologically transitive}; that is, for any two subintervals $U_1,U_2\subseteq I$, there exist $x_0\in U_1$ and an integer $n>0$ such that $f^{n}(x_0)\in U_2$.
    
    \item $f$ exhibits sensitive dependence on initial conditions; namely, there exists a constant $\delta>0$ such that, for every $x_0\in I$ and every open interval $U$ containing $x_0$, there exist $y_0\in U$ and an integer $n>0$ satisfying $\left|f^{n}(x_0)-f^{n}(y_0)\right|>\delta$.
    
\end{enumerate}
Therefore, an invariant set that attracts the nearby trajectories and is chaotic is known as a \textit{chaotic attractor}. Let $X$ be a topological space and $f:X\rightarrow X$ be a map. If $p$ is a fixed point of $f$, the \emph{stable} and \emph{unstable} sets of $p$ are defined as
\[
W^{s}(f,p)=\{q\in X:\; f^{n}(q)\rightarrow p \text{ as } n\rightarrow\infty\},
\]
and
\[
W^{u}(f,p)=\{q\in X:\; f^{-n}(q)\rightarrow p \text{ as } n\rightarrow\infty\},
\]
where $f^{-1}$ denotes the inverse of $f$. If $p$ is a periodic point with least period $k$, then $p$ is a fixed point of $f^k$, and
\[
W^{s}(f,p)=W^{s}(f^{k},p), \qquad
W^{u}(f,p)=W^{u}(f^{k},p).
\]
Crisis refers to a sudden qualitative change of chaotic dynamics such as change in size, appearance or destruction of chaotic attractor \cite{PhysRevLett.48.1507}. These changes occur due to the collision of the chaotic attractor with an unstable invariant set, typically, a collision with the stable manifold of the unstable periodic orbit. In general, crises are broadly distinguished into three major types: boundary crises, in which a chaotic attractor is suddenly created or destroyed, interior crises, in which a chaotic attractor experiences a sudden change in size and shape, and attractor merging crises, in which a number of symmetrically disposed chaotic attractors merge (or, inversely, split).

A bifurcation diagram is a fundamental tool for visualizing qualitative changes in the chaotic behavior of a dynamical system as a control parameter is varied. For discrete-time systems (maps) of the form $\mathbf{x}_{t+1}=\mathbf{f}(\mathbf{x}_t,\mu)$, the state variables are iteratively computed for each value of $\mu$. After discarding an initial transient, the remaining asymptotic states are plotted directly on the vertical axis with respect to the bifurcation parameter. In contrast, for continuous-time systems governed by ordinary differential equations, $\dot{\mathbf{x}}=\mathbf{f}(\mathbf{x},\mu)$, plotting the entire time series produces a dense, uninformative band. Instead, after numerical integration and transient removal, only the local extrema of a selected state variable $x_i(t)$; either maxima ($\dot{x}_i=0,\ \ddot{x}_i<0$) or minima ($\dot{x}_i=0,\ \ddot{x}_i>0$) are recorded on the vertical axis. By plotting asymptotic states for maps and local extrema for flows against the control parameter, bifurcation diagrams provide a unified representation that clearly reveals period-doubling cascades, chaotic bands, periodic windows, and other dynamical transitions.

\subsection{Scaling of transient times}\label{subsec:scaling}
Beyond a boundary crisis, trajectories are no longer permanently confined to the  chaotic attractor preceding the crisis. Instead, they remain for a finite duration in the remnant of the destroyed attractor before eventually escaping. This interval is referred to as the chaotic transient. Although the transient lifetime depends strongly on the initial condition, its ensemble average obeys a universal scaling law for boundary crisis. The average length of chaotic transient $\langle\tau\rangle$ scales as: $\langle\tau\rangle = (\mu - \mu_c)^{-\gamma}$ where $\mu_c$ is the bifurcation point and $\gamma$ is the scaling exponent. For the logistic map, the exponent is known to be $\gamma=1/2$. An analytical derivation of this exponent can be obtained for one-dimensional maps of the form $x_{t+1}=f(x_t,\mu)$. Let $x_e$ denote the escape time satisfying $f(x_e,\mu_c)=X_{boundary}$, where $X_{boundary}$ is the boundary of the chaotic interval beyond which trajectories rapidly approach a different attractor. In the logistic map, $x_e=0.5$ and $X_{boundary}=1$, since for $\mu>\mu_c=4$, any iterate with $x>X_{boundary}$ becomes negative under subsequent iterations and diverges to $-\infty$.
To determine the scaling exponent \cite{Ott2002_8}, the bifurcation parameter is written as $\mu=\mu_c+\epsilon$ and the state variable is expanded about the escape point as $\delta x=x-x_e$. A Taylor expansion of the map yields:
\begin{equation}
    \begin{aligned}
        f(x_e + \delta x, \mu_c + \epsilon) =& f(x_e, \mu_c) \;+\;  \frac{\partial f}{\partial x}\delta x \; + \; \frac{\partial f}{\partial \mu} \epsilon \\&+ \frac{1}{2} \frac{\partial^2 f}{\partial x^2} (\delta x)^2 + higher \; order. \nonumber
    \end{aligned}
\end{equation}
All the partial derivatives are evaluated at $x_e, \mu_c$. Since $f(x_e, \mu_c) = X_{boundary}$ and for escape $f(x, \mu) > X_{boundary}$, the escape condition becomes the following:
\begin{equation}
    \begin{aligned}
        \frac{\partial f}{\partial x}\delta x \; + \; \frac{\partial f}{\partial \mu} \epsilon \;+\; \frac{1}{2} \frac{\partial^2 f}{\partial x^2} (\delta x)^2 > 0. \nonumber
    \end{aligned}
\end{equation}
This can be expressed in the form $A\epsilon + B (\delta x)^2 + C \delta x > 0$. For the logistic map, the quadratic nonlinearity places the maximum of the return map ($x_{t+1}$ vs $x_t$) exactly at the escape point, implying $B<0$ and $C=0$. The escape condition therefore reduces to $(\delta x)^2 < -A\epsilon / B $ or $\delta x \sim \sqrt{\epsilon}$. Thus, whenever the trajectory enters an interval of width $\delta x$ surrounding the escape point, it escapes the chaotic region and the transient terminates. The invariant density $\rho(x) dx$ is defined as the probability that a point on a long trajectory will lie between $x$ and $x + dx$. Numerically, this density can be estimated by constructing a histogram of iterates on the chaotic attractor. For the logistic map, the invariant density is uniform in the vicinity of the escape region, implying that the probability of entering the interval of width $\delta x$ is proportional to $\delta x$, and hence to $\sqrt{\epsilon}$. The transient length is inversely proportional to this probability, therefore $\langle\tau\rangle = (\mu - \mu_c)^{-1/2}$. 

\subsection{Delay coordinate embedding}\label{subsec:delay}
Let $M$ and $N$ be smooth manifolds, and let $f:M\rightarrow N$ be a smooth map. In differential topology, $f$ is called a smooth embedding if it is a homeomorphism onto its image and its derivative $Df_p:T_pM\rightarrow T_{f(p)}N$ is injective for every $p\in M$. Whitney's weak embedding theorem \cite{whitney1936differentiable} states that if $M$ is a smooth manifold of dimension $m$, then there exists a smooth embedding $f : M \rightarrow \mathbf{R}^{2m+1}$. Takens' embedding theorem \cite{takens1981detecting, PhysRevLett.45.712} establishes criteria under which the partial observations of a dynamical system can be used to reconstruct the full state. This reconstruction is possible because the original trajectory and the time-lags of the partially observed trajectory are related through an embedding.
Specifically, for a deterministic dynamical system evolving on a $d$-dimensional invariant manifold (such as a chaotic attractor), suppose we measure only a single scalar observable, $y(t) = h(\mathbf{x}(t))$, where $h$ is a generic smooth measurement function. Takens proved that the delay coordinate map $\Phi$, which constructs an $m$-dimensional state vector from sequentially delayed measurements given by $\Phi(\mathbf{x}) = [y(t), y(t-\tau), y(t-2\tau), \dots, y(t-(m-1)\tau)]$, with $\tau$ being a chosen time delay; generically forms a smooth embedding. Crucially, as long as the embedding dimension satisfies the condition $m \ge 2d + 1$, a faithful reconstruction is possible. Because of this topological equivalence, the invariant dynamical properties of the true system, such as Lyapunov exponents, correlation dimensions, and topological entropy, are perfectly preserved in the reconstruction space, enabling the analysis of complex multidimensional dynamics from a single univariate time series.
The quality of the reconstructed attractor is highly sensitive to the choice of the time delay $\tau$, and several frameworks for its systematic selection have been proposed \cite{ROSENSTEIN199482,tan2023selecting}.

\subsection{Finite time Lyapunov exponents}\label{subsec:ftle}
For a $d$-dimensional dynamical system, the Lyapunov spectrum consists of $d$ exponents that quantify the average exponential rates at which infinitesimal perturbations grow or decay along different directions in phase space. Among these, the largest Lyapunov exponent characterizes the average divergence of nearby trajectories and is defined as \cite{Ott2002_9}
\begin{equation}
    \Lambda_{max} = \lim_{t \to \infty} \frac{1}{t} ln \frac{D(t)}{D(0)}, \nonumber
\end{equation}
where $D(0)$ denotes the infinitesimal initial separation between two trajectories and $D(t)$ is their separation after a time $t$. Since this definition involves the limit $t\rightarrow\infty$, it represents an asymptotic measure of the system's stability. To characterize the local stretching behavior over finite time intervals, finite-time Lyapunov exponents (FTLEs) are employed \cite{PhysRevE.60.2761}. The trajectory is partitioned into segments of length $N$, and the FTLE associated with the $k^{\mathrm{th}}$ segment is computed as:
\begin{equation} \label{eq3}
    \lambda_N (k) = \frac{1}{N} \sum_{j = N(k-1) + 1}^{kN} \lambda_{max} (j),
\end{equation}
where $\lambda_{max} (j)$ is  calculated for a single step. To compute $\lambda_N$, we initialize the system with a reference trajectory and an infinitesimally perturbed copy. In practice, a reference trajectory and an infinitesimally perturbed counterpart are evolved simultaneously, and the instantaneous stretching rates are recorded at every iteration. The $N$-step FTLE is then obtained by averaging these single-step exponents over a moving window of length $N$, as described by Eq.~(\ref{eq3}). Consecutive windows overlap by $N-1$ time steps, producing an ensemble of FTLE values. Because FTLEs are local quantities, they depend on the initial condition of the trajectory segment. Let $P(\lambda,N) d\lambda$ denote the probability that the $N$-step FTLE lies within the interval $[\lambda,\lambda+d\lambda]$. The resulting distribution exhibits characteristic features that reflect the underlying dynamical properties of the attractor \cite{PhysRevE.60.2761}. As the averaging time increases, the FTLE distribution progressively narrows, and in the limit of large $N$, it converges to the asymptotic Lyapunov exponent ($\lim_{N \to \infty} P(\lambda, N) \to \delta(\Lambda - \lambda)$).

\section{Dynamical insight to successful Predictions}
\begin{figure}[t!]
    \includegraphics[width=0.43\textwidth]{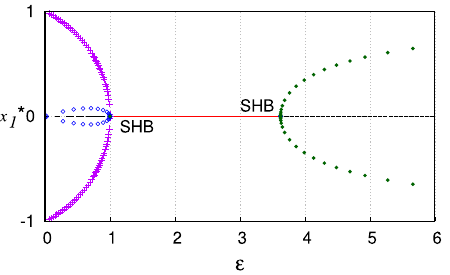}
    \caption{Reused from \cite{PhysRevE.110.034211}. Bifurcation diagram of the coupled Stuart-Landau system. Solid green and hollow blue circles denote stable and unstable limit cycles, respectively, while solid red and dashed black lines represent stable and unstable fixed points, respectively. Purple crosses indicate the torus branch, and SHB denotes the supercritical Hopf bifurcation.}
    \label{fig2}
\end{figure}
During the testing phase, the output of the current state acts as the input of the next state. Using this information, we can modify Eq.~\ref{eq1} to obtain:
\begin{equation}
\begin{aligned}
    r[i+1] =& (1-\alpha)r[i]  +  \alpha \hspace{0.1cm} tanh(A r[i] + \\&W_{in}W_{out} r[i] + k_{b} W_b (\varepsilon - \varepsilon_b)), \nonumber
\end{aligned}
\end{equation}
which can be further simplified to a simple map equation;
\begin{equation} \label{eq4}
    r[i+1] = (1-\alpha)r[i]  +  \alpha \hspace{0.1cm} tanh(\Lambda r[i] + \Omega).
\end{equation}
This forms a system of $m$ autonomous dynamical equations, where $m$ is the dimension of the reservoir, $r[i]$ and $\Omega$ both are $m$ dimensional column vectors, while $\Lambda$ is a $m \times m$ matrix. The matrix $\Lambda$ is ($\mathcal{A} + W_{in}W_{out}$) and $\Omega$ equals ($k_b W_b (\varepsilon - \varepsilon_b)$). Consequently, irrespective of whether the original dynamical system is continuous or discrete-time, the trained reservoir evolves as a discrete-time map. This unified representation allows the learned reservoir dynamics to be analyzed using the nonlinear dynamical tools developed for maps, introduced in the previous section.

\subsection{Prototype models}
\begin{figure}[t!]
    \includegraphics[width = \linewidth]{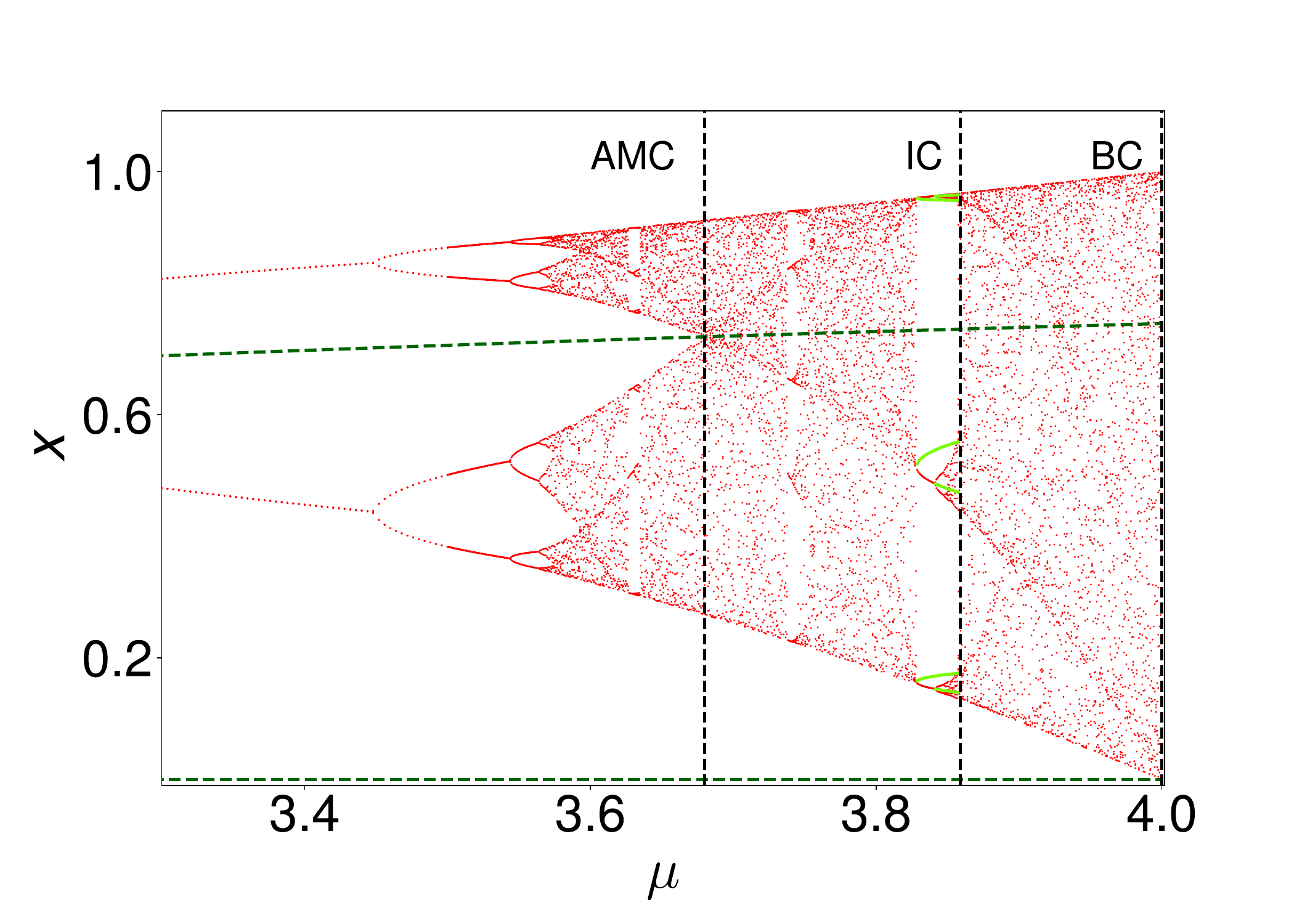}
    \caption{Crises in a logistic map. The dashed dark green curves represent the two fixed points, $x=0$ and $x=1-1/\mu$, while the light green curve denotes the unstable period-3 orbit. The three vertical dashed black lines mark the parameter values corresponding to different crisis events: attractor-merging crisis (AMC) at $\mu=3.68$, interior crisis (IC) at $\mu=3.86$, and boundary crisis (BC) at $\mu=4$.}
    \label{fig3}
\end{figure}
A diffusively coupled, non-identical Stuart-Landau oscillators \cite{ARONSON1990403} is governed by
\begin{eqnarray}
    &&\dot x_{i} = (1 - x_{i}^{2} - y_{i}^{2}) x_{i} - \omega_{i} y_{i} + \varepsilon (x_j - x_i),\nonumber\\
    &&\dot y_{i} = (1 - x_{i}^{2} - y_{i}^{2}) y_{i} + \omega_{i} x_{i} + \varepsilon (y_j - y_i), \nonumber
\end{eqnarray}
where $\omega_1=2$ and $\omega_2=7$. The bifurcation structure, computed using the XPPAUT package \cite{10.1115/1.1579454}, is shown in Fig.~\ref{fig2}. The system exhibits two supercritical Hopf bifurcations. For $1<\varepsilon<3.6$, the stable attractor is a fixed point. At $\varepsilon\approx3.6$, this fixed point loses stability through a Hopf bifurcation, giving rise to a stable limit cycle that persists for $\varepsilon>3.6$. A second Hopf bifurcation occurs near $\varepsilon\approx1$, where the phase-space dynamics differs qualitatively. As $\varepsilon$ is decreased below this value, the fixed point again loses stability, but the emerging limit cycle immediately undergoes a torus bifurcation, producing a stable torus. Consequently, the attractor for $\varepsilon<1$ is a torus, in contrast to the simple closed invariant curve associated with the stable limit cycle observed for $\varepsilon>3.6$.

The logistic map is described by the recursive relation $x_{t+1} = \mu x_t(1- x_t)$, where $x \in [0, 1]$. It is a minimal model that can exhibit a broad spectrum of complex behaviors, including fully developed chaos, intermittency, and crisis induced transitions. As shown in Fig. \ref{fig3}, at $\mu = 4$, the chaotic attractor collides with the unstable fixed point $x = 0$ beyond which all the trajectories initialized at different initial conditions eventually diverge to $-\infty$, causing a boundary crisis.
\begin{figure}[t!]
\includegraphics[width=0.46\textwidth]{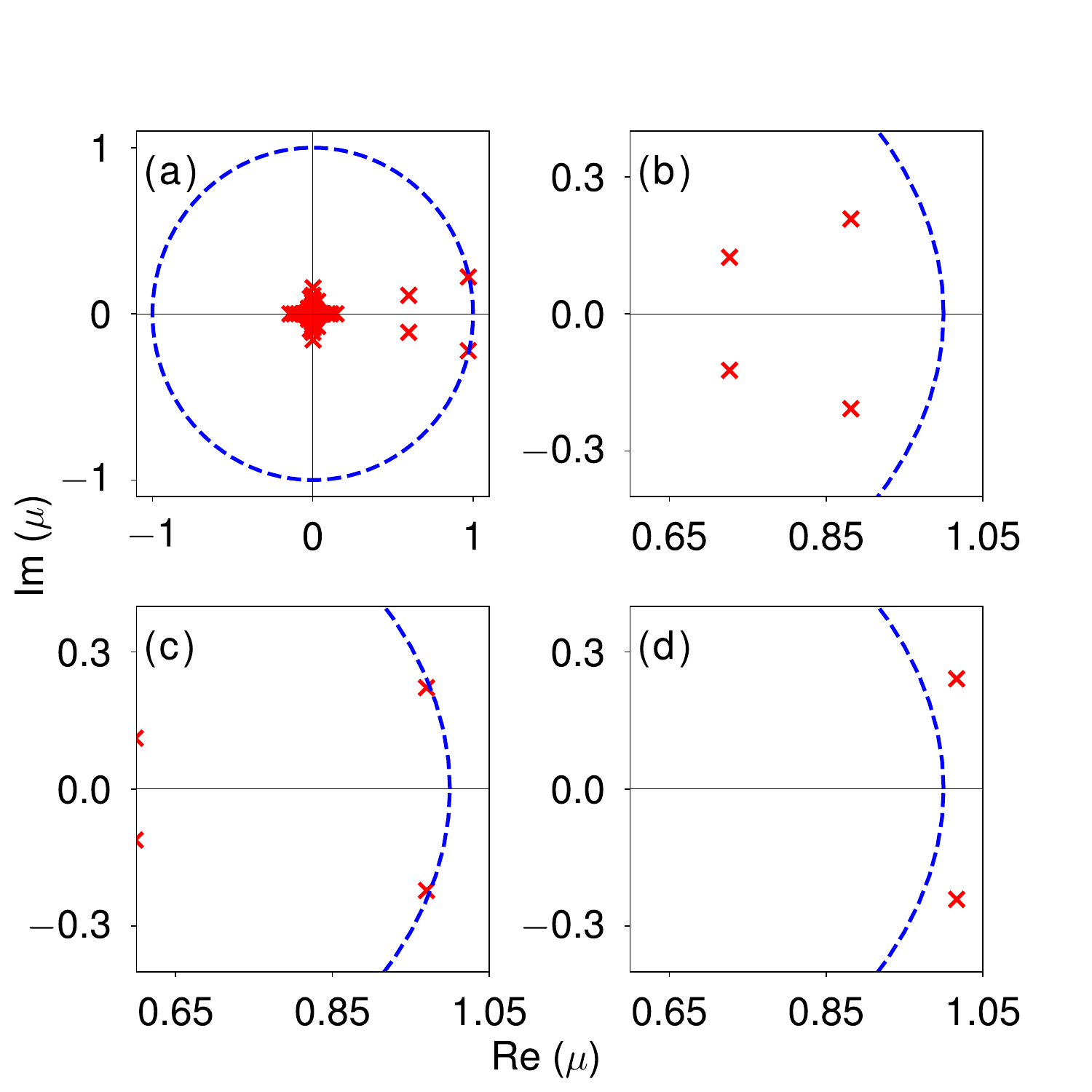}
    \caption{Reproduced from \cite{PhysRevE.110.034211}. Evolution of the eigenvalues ($\mu$) of the learned reservoir map for the coupled Stuart-Landau system as the bifurcation parameter $\varepsilon$ is varied across the limit-cycle transition. The blue dashed curve denotes the unit circle in the complex plane, while red crosses indicate the eigenvalues of the Jacobian $\mathcal{J}$. (a) A single complex conjugate pair crosses the unit circle at $\varepsilon=3.62$. Panels (b)-(d) show the corresponding eigenvalue distributions at $\varepsilon=3.35$, $3.62$, and $3.90$, respectively.}
    \label{fig4}
\end{figure}
The merging of attractors can be seen in the interval $3.57 < \mu <3.68$. Starting from the onset of chaos ($\mu$ = 3.57), at each such crisis the number of pieces of chaotic attractor is halved \cite{ananthkrishnan2001crises}. The most prominent attractor merging crisis happens at $\mu = 3.68$ where two pieces merge to yield a single piece chaotic attractor. At each period doubling bifurcation, an unstable periodic orbit is created. These orbits collide with the pieces of chaotic attractor causing a cascade of ``halving" of the number of pieces of the attractor. The final crisis point is the collision of unstable periodic orbit of period 1 (fixed point: $x^* = 1- 1/\mu$) with the two pieces of chaotic attractor.
The interior crisis, is characterized by a sudden expansion of the chaotic attractor \cite{PhysRevLett.48.1507}. In the logistic map, such an interior crisis occurs at $\mu \approx 3.8568007$, where a collision between an unstable period-3 orbit and the chaotic attractor leads to an abrupt increase in attractor size.
\begin{figure}[t!]
    \begingroup
    \begin{tabular}{c}
    \includegraphics[width = 0.23\textwidth,height=4cm]{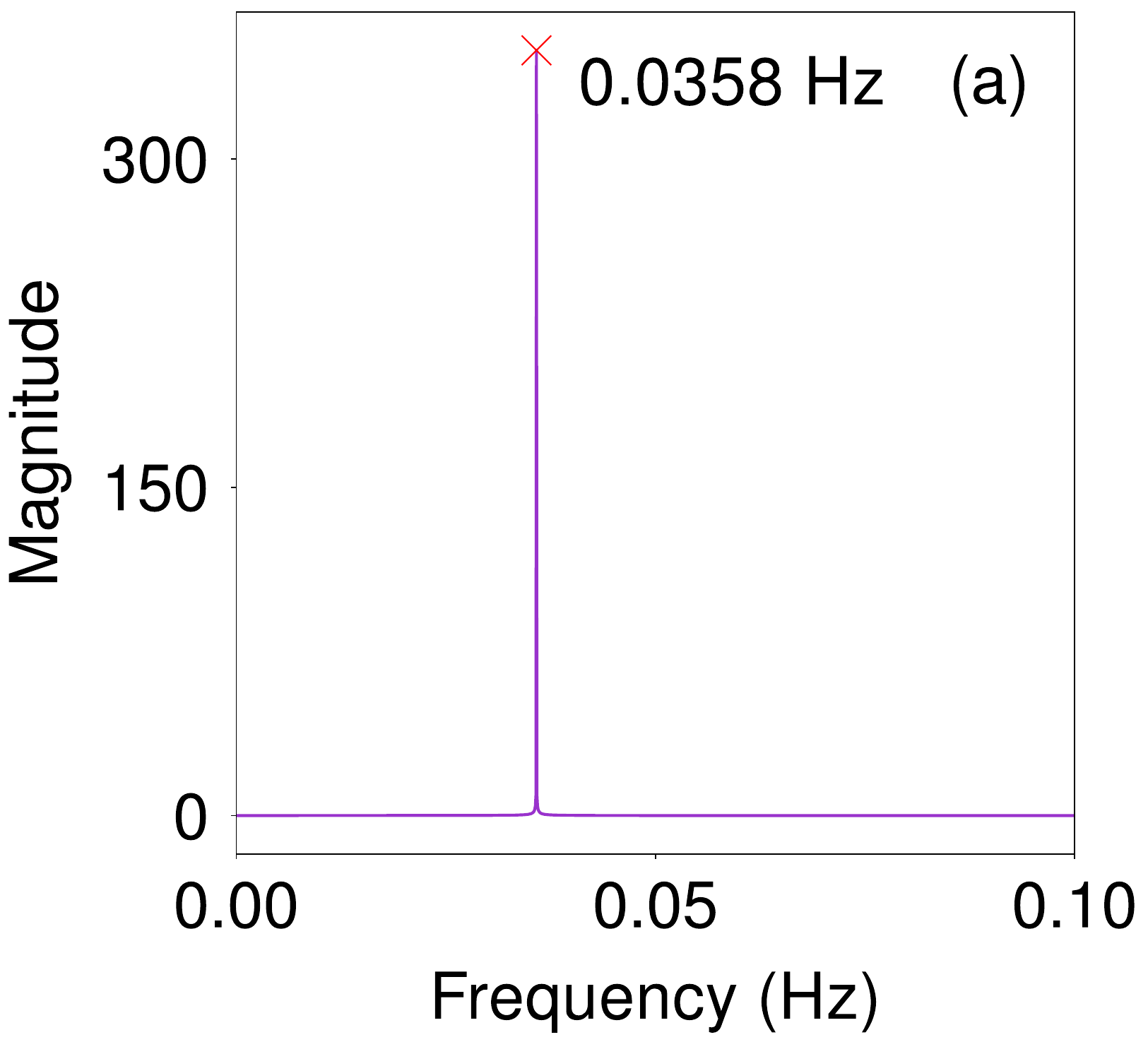}
    \includegraphics[width = 0.23\textwidth,height=4cm]{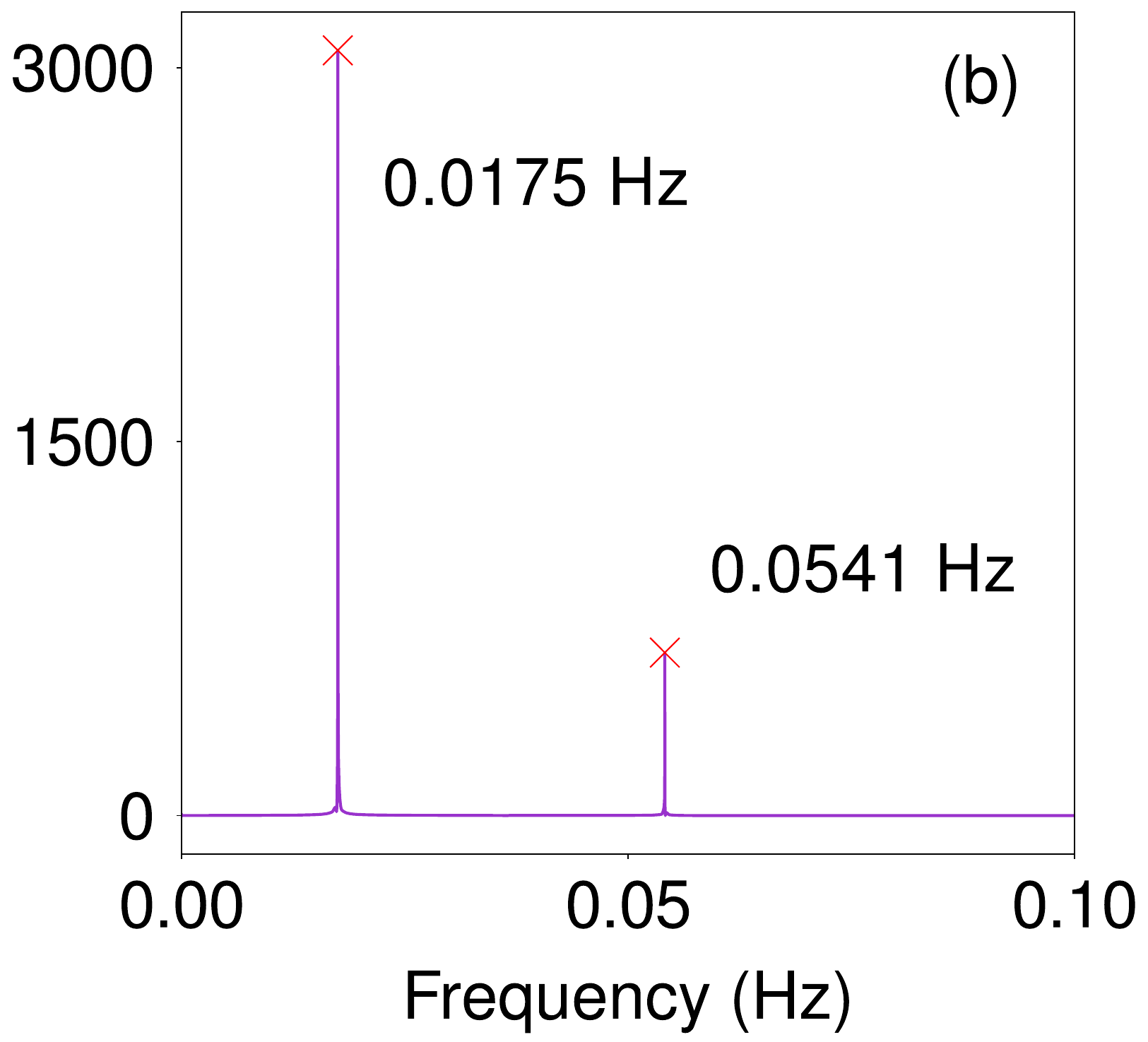}
    
    \end{tabular}
    \endgroup
    \caption{Reproduced from \cite{PhysRevE.110.034211}. The dominant frequencies of the time series are isolated along with their magnitudes by performing the Fourier transform of $x_1(t)$ for coupled Stuart-Landau oscillator at (a) $\varepsilon = 3.63$, (b) $\varepsilon = 0.99$. }
    \label{fig5}
\end{figure}
\subsection{Analysis of Jacobian Eigenvalues for Characterizing Bifurcation Phenomena}\label{subsec:RCjac}
A successful prediction implies that the trained output matrix $W_{out}$ enables the reservoir map in Eq.~\ref{eq4} to faithfully reproduce the bifurcation structure of the original continuous dynamical system. Since the predicted output is obtained as a linear projection of the reservoir state, i.e., $v[i+1]=W_{out}r[i]$, a thorough study of Eq.~\ref{eq4} will explain the final output. The fixed point $r^{*}$ of Eq.~\ref{eq4} is first computed numerically after which the Jacobian matrix $\mathcal{J}$ is evaluated at $r^*$ as follows:
\begin{equation}
\begin{aligned}
    \delta r[i+1] &= \big((1-\alpha) + \{\mathbb{I} - tanh^{2}(\Lambda r^{*} + \Omega)\}\Lambda\big) \delta r[i] \\&
    = \mathcal{J} \delta r[i].
\end{aligned} \nonumber
\end{equation}
The evolution of the eigenvalues of $\mathcal{J}$ is then examined as the bifurcation parameter $\varepsilon$ is varied \cite{PhysRevE.110.034211}. 

For the coupled Stuart-Landau system, analysis is restricted to the neighborhood of the Hopf bifurcation ($\varepsilon = 3.6$). The reservoir hyperparameters are chosen as $(m=300, b=c=2, \alpha=1, \rho=0.18, \sigma=0.03, k_{b}=0.52, \varepsilon_b=0)$, and the RC is trained using data at $\varepsilon=3.75$, $3.70$ and $3.65$. After training, the reservoir predicts the bifurcation at $\varepsilon\approx3.62$, which closely matches the bifurcation point of the original system. To understand this behavior. The evolution of the eigenvalues of $\mathcal{J}$ with the bifurcation parameter $\varepsilon$ is then examined as depicted in Fig.~\ref{fig4}. The spectrum consists of both real eigenvalues and complex conjugate pairs. For $\varepsilon<3.62$, all eigenvalues lie strictly inside the unit circle, indicating that the fixed point of the learned map is stable. Consequently, the reservoir states converge to a stable spiral, and since the predicted variables ($x_i,y_i$) are linear combinations of these states, they also exhibit stable spiral dynamics, consistent with the amplitude-death regime of the original system.
\begin{figure}[t!]
    
    \includegraphics[width=0.46\textwidth]{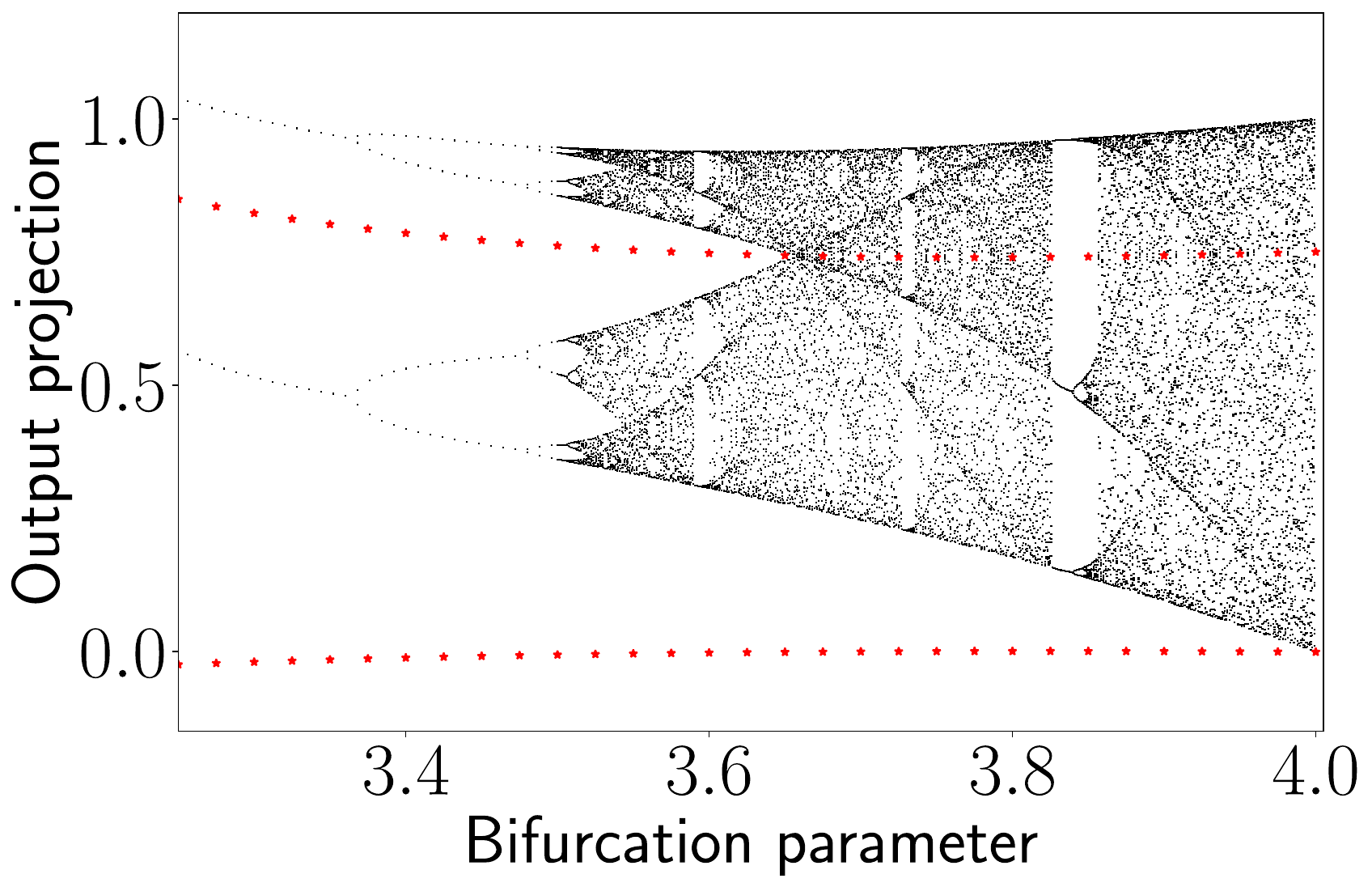}

    \caption{Reproduced from \cite{8b7z-fnjd}. Crises in trained RC map due to collision of two unstable fixed points with the chaotic attractor. Collision close to $\mu = 3.68$ is responsible for attractor merging crisis and the collision close to $\mu = 4$ causes boundary crisis.}
    \label{fig6}
\end{figure}
As $\varepsilon$ increases beyond $3.62$, a single complex conjugate pair exits the unit circle, while all remaining eigenvalues remain inside it (Fig.~\ref{fig4}). This crossing signifies a Neimark-Sacker bifurcation of the learned map, corresponding to the transition from amplitude death to sustained oscillations in the original system. Beyond the bifurcation, all $m=300$ reservoir variables evolve with a common oscillation period, generating invariant closed curves in every two-dimensional projection $(r_j,r_k)$ of the reservoir phase space. Because the output variables are linear combinations of the reservoir states, the predicted time series $(x_i,y_i)$ inherit the same oscillatory behavior. Thus, the learned map naturally reproduces the stable limit-cycle dynamics of the original system beyond the bifurcation point.

After successful training in the vicinity of $\varepsilon = 1$, the eigenvalue analysis was repeated. Unlike the case of a single conjugate pair crossing, observed near $\varepsilon = 3.6$, the generated map corresponding to $\varepsilon = 1$ exhibits two conjugate pair crossings. A single pair crossing gives rise to a closed invariant curve, as observed for $\varepsilon = 3.6$. Consequently, accurate prediction of the torus requires the emergence of a second conjugate pair crossing in the learned map. Torus formation has previously been associated with a double Neimark-Sacker bifurcation (a codimension-2 bifurcation), where two conjugate pairs of eigenvalues simultaneously cross the unit circle \cite{LUO2006154,article}. However, the reservoir map (Eq. \ref{eq4}) undergoes a codimension-1 bifurcation (involving only a single bifurcation parameter), implying that at most one conjugate pair can cross the unit circle at the bifurcation point \cite{Kuznetsov2023}. Therefore, the first crossing occurs through a conventional Neimark-Sacker bifurcation, followed by the crossing of a second conjugate pair \cite{PhysRevE.110.034211}.
\begin{figure}[t!]
     \includegraphics[scale=0.3]{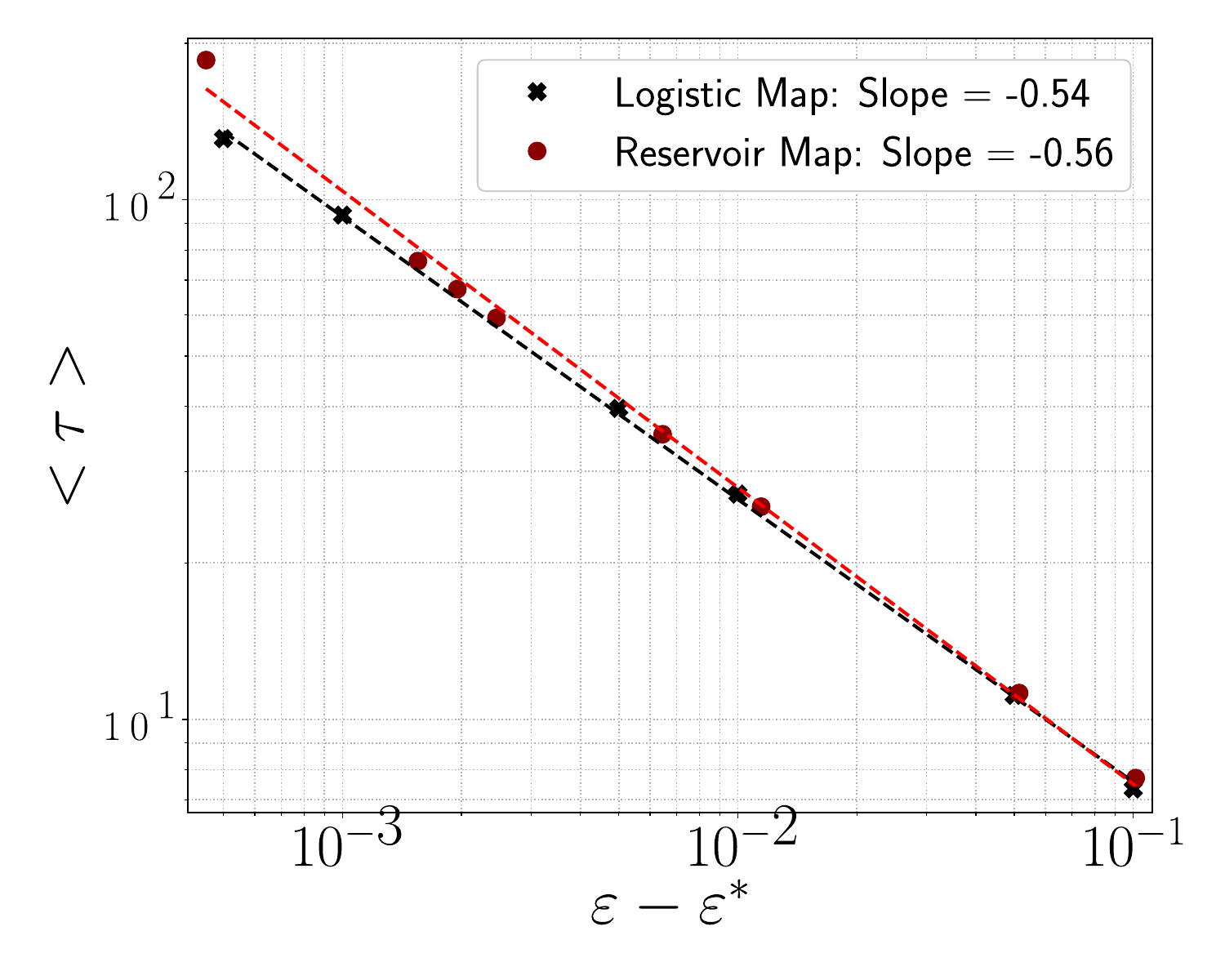}
    \caption{Reproduced from \cite{8b7z-fnjd}. $\langle \tau \rangle$ is plotted as a function of $\varepsilon - \varepsilon^*$ ($\mu - \mu_c$ for the logistic map). Trained reservoir map matches the scaling exponent of the logistic map. The average over $\tau$ is performed using 10000 different initial conditions.}
    \label{fig7}
\end{figure}
The frequency content of all the four time series ($x_1(t), x_2(t), y_1(t), y_2(t)$) is examined using Fourier analysis for both the dynamical regimes (Fig.~\ref{fig5}). In the limit-cycle regime ($\varepsilon = 3.6$), the dynamics are characterized by a single fundamental frequency common to all four time series. By contrast, the torus regime exhibits two distinct frequencies in each time series. Although different random realizations of $W_b$, $W_{in}$, and $\mathcal{A}$ lead to different learned maps for both $\varepsilon = 3.6$ and $\varepsilon = 1$, accurate reconstruction of the dynamics requires the frequency spectrum of the generated time series to remain unchanged. This consistency is ensured by preserving the arguments of the eigenvalues crossing the unit circle, where $\theta = tan^{-1}(Im(\mu)/Re(\mu))$. Consequently, irrespective of the particular random realization, the crossing eigenvalues occupy the same angular positions with respect to the real axis in the Argand plane. For the limit-cycle case ($\varepsilon = 3.6$), let $\mu = e^{\pm i\theta(0)}$ represent the conjugate eigenvalue pair crossing the unit circle. The corresponding value, $\theta(0)/2\pi = 0.0358$, is in close agreement with the dominant frequency extracted from the Fourier spectrum of the original system. At the torus bifurcation ($\varepsilon = 1$), the crossing eigenvalue pairs are denoted by $\mu_1 = e^{\pm i\theta_1}$ and $\mu_2 = e^{\pm i\theta_2}$, with $\theta_1/2\pi = 0.0174$ and $\theta_2/2\pi = 0.0541$. Both values closely coincide with the two characteristic frequencies obtained from the Fourier spectra of the original system time series.

\subsection{Projection of high-dimensional RC manifold onto low-dimensions}\label{subsec:RCCrisis}
The RC is trained using chaotic time series generated by the logistic map ($x_{t+1} = \mu x_t (1-x_t)$) for $\mu \in {3.92, 3.93, 3.94\; \text{and}\;  3.95}$. The corresponding hyperparameters are chosen as $m = 400, \alpha = 0.86, b = 2.13, c = 1.15, \sigma = 0.526, \rho = 0.9, k_b = 1$, and $\varepsilon_b = 0.3$. After training, the RC-generated time series successfully reproduces the discontinuous transition, predicting the critical point at approximately $\mu_c \sim 3.9985$. To identify the dynamical mechanism responsible for this prediction, the fixed points of the learned reservoir map are computed and their stability is examined. Owing to the high-dimensional nonlinear nature of the reservoir architecture, identifying all fixed points using conventional approaches, such as local searches initialized from random seeds, becomes computationally challenging. An alternative framework based on \textit{Directional Fibers} \cite{8016349,directional-fibers}, specifically developed for RNN architectures, provides an efficient means of locating a large number of fixed points while remaining computationally competitive. The stability of each fixed point is subsequently determined by evaluating the maximum eigenvalue of the Jacobian of the reservoir map \cite{8b7z-fnjd}. The analysis reveals multiple fixed points, of which only two are stable, while the remaining ones are unstable. Beyond $\mu = 4$, these stable fixed points constitute the only attractors of the learned reservoir dynamics. As a result, trajectories converge to these attractors instead of diverging to $-\infty$, as occurs in the original logistic map.

Direct interpretation of crisis mechanisms within the high-dimensional reservoir state space is inherently difficult. To facilitate analysis, the reservoir dynamics are projected onto a one-dimensional space through the output matrix $W_{out}$. The bifurcation diagram of the learned map (Fig.~\ref{fig6}) is obtained by mapping Eq.~\ref{eq4} back to one dimension using $W_{out}$, while the corresponding fixed points computed in the reservoir space are projected in the same manner. Near $\mu \sim 4$, an unstable fixed point collides with the chaotic attractor, causing its disappearance. Around this parameter value, the unstable fixed point approximates the original fixed point $x^* = 0$, thereby confirming that the learned reservoir map reproduces the boundary crisis mechanism.
The one-dimensional projection of the learned reservoir map also captures the attractor-merging crisis. At $\mu = 3.64$, two previously disconnected chaotic attractors merge into a single chaotic attractor exactly when an unstable fixed point (corresponding to a period-1 unstable periodic orbit) collides with them (Fig.~\ref{fig6}). This unstable fixed point persists for larger values of $\mu$ and closely follows the analytical fixed point $x^* = 1-1/\mu$. These observations demonstrate that the trained reservoir map successfully learns and reproduces the attractor-merging crisis in its one-dimensional projection.
Previous studies have demonstrated that trained reservoir computers can accurately reproduce statistical characteristics including the scaling exponent of transient lifetimes \cite{Kong_2021}. Although averaging over multiple independently initialized reservoirs can recover this scaling behavior, each random initialization modifies the underlying high-dimensional reservoir dynamics, making it difficult to examine how an individual trained reservoir captures the correct scaling.
Therefore, to understand how the scaling exponent of trained RC matches the true system, a different approach for scaling exponent calculation was applied \cite{8b7z-fnjd}. For the case of trained reservoir map, the 1-d projection is as follows:
\begin{equation}
\begin{aligned}
    x_{i+1} =& W_{out} \Big ((1-\alpha) r_{i} \\&+  \alpha \;tanh(Ar_{i} + W_{in}x_{i} + \Omega(\varepsilon - \varepsilon_b)) \Big). \nonumber
\end{aligned}
\end{equation}
The matrices $A$, $W_{in}$, and $\Omega$ are randomly initialized only once and remain fixed throughout the training process. After training, the output matrix $W_{out}$ is also fixed, ensuring that the learned reservoir map remains identical during the entire scaling analysis. The computation begins by selecting random initial conditions $x_0$ from the interval $(0,1)$. The logistic map is first iterated to generate a warm-up sequence that initializes the reservoir state. This reservoir state is then evolved autonomously at a fixed bifurcation parameter until the condition $u[i+1] < 0$ is satisfied. The number of iterations required to reach this threshold is recorded for each initial condition, and the average over all realizations yields the mean transient lifetime $\langle\tau\rangle$. The value of the scaling exponent is governed by the shape of the return map $(x_{t+1}\; vs \; x_t)$ of the trained reservoir together with the natural invariant density $\rho(x)dx$. The one-dimensional projection of the trained reservoir exhibits the same parabolic return map, with a single maximum located at the escape point $x=0.5$, while its invariant density closely matches that of the logistic map. This agreement provides a dynamical explanation for the reservoir's ability to accurately reproduce the scaling exponent of the original logistic map (Fig.~\ref{fig7}).

\subsection{FTLE distributions for learning crisis}\label{subsec:RCFtle}
Beyond the boundary and attractor-merging crises, extending the same methodology to an interior crisis is considerably more challenging. The high-dimensional nature of the reservoir map makes the explicit computation of unstable periodic orbits and the identification of their collisions with the chaotic attractor computationally infeasible. Consequently, the dynamical mechanism responsible for the interior crisis cannot be verified directly using the approach employed for boundary and attractor-merging crises. Although the bifurcation diagram of the trained reservoir map (Fig.~\ref{fig6}) exhibits close agreement with that of the logistic map, such similarity alone does not guarantee that both systems undergo identical transition mechanisms, particularly in view of the high dimensionality of the reservoir dynamics.
\begin{figure}[t!]
    \centering
    \begin{minipage}{0.49\linewidth}
        \centering
        \includegraphics[width=\linewidth]{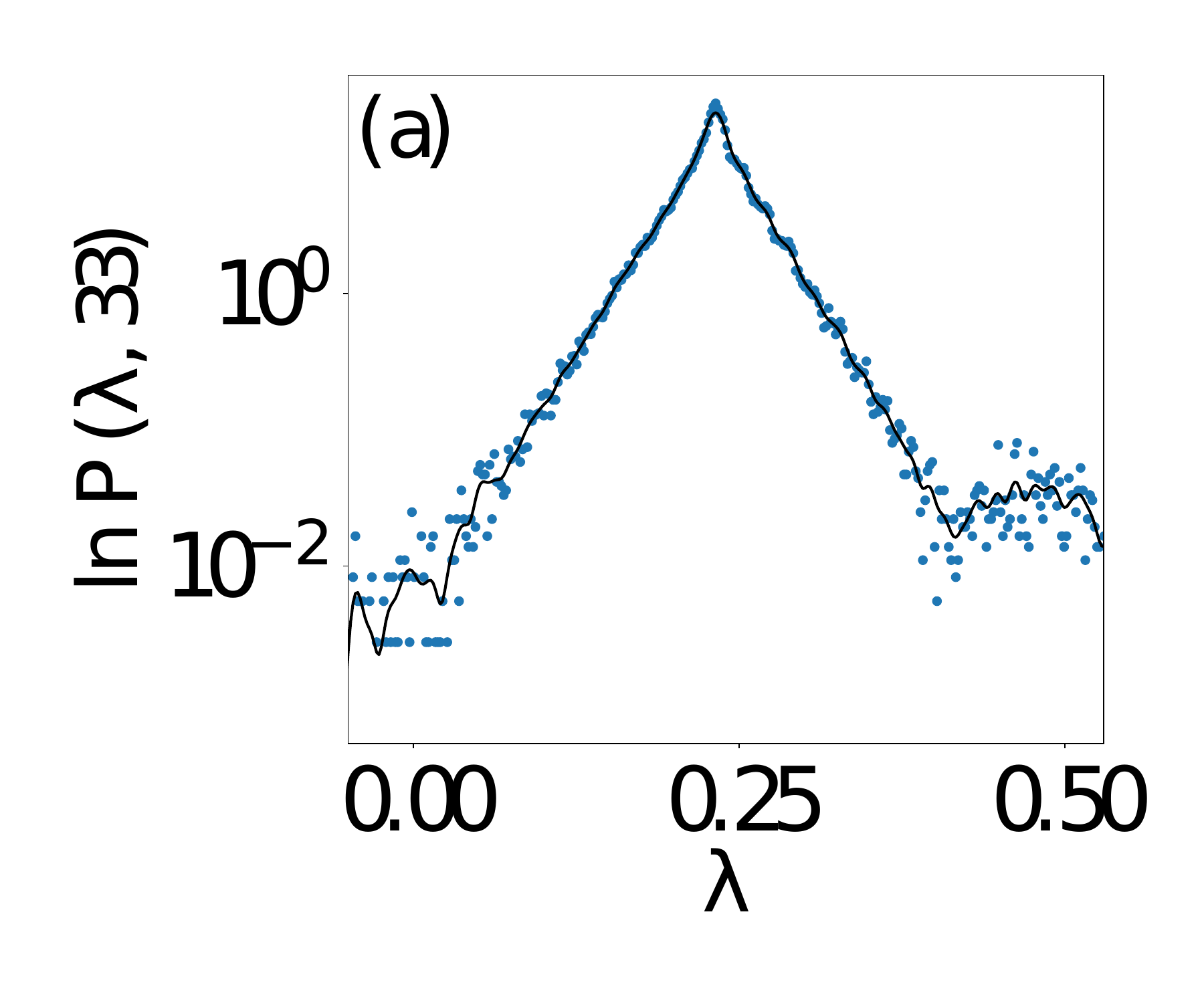}
    \end{minipage}
    \hfill
    \begin{minipage}{0.49\linewidth}
        \centering
        \includegraphics[width=\linewidth]{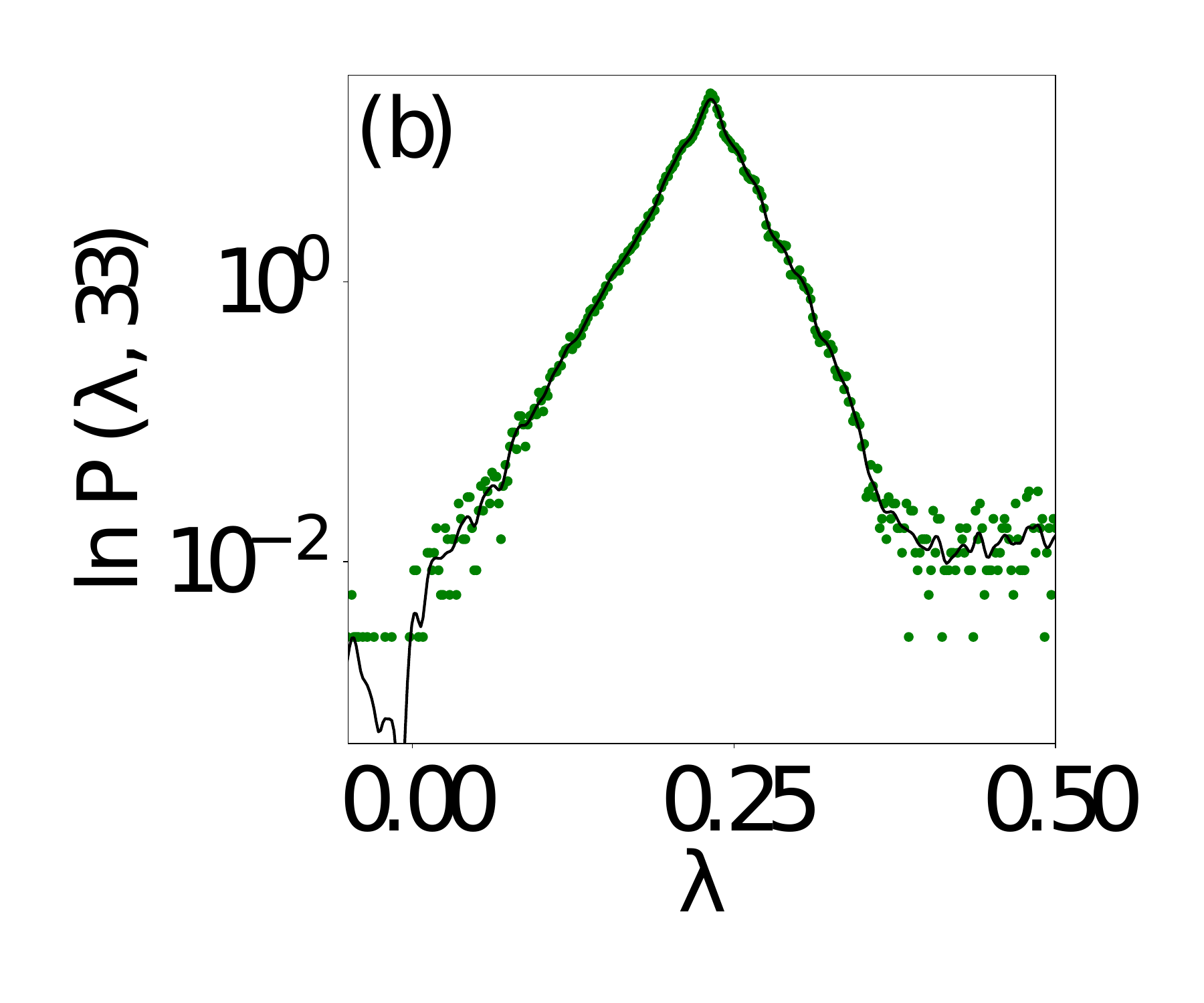}
    \end{minipage}
    \caption{Reproduced from \cite{sisodia2026finite}. Probability density of finite-time Lyapunov exponents at interior crisis computed from segments of length $N=33$ using $200000$ iterations. 
    (a) Logistic map at $\mu = 3.8568007$; the solid black curve denotes the kernel density estimate for the smooth approximation of the probability density function. 
    (b) FTLE distribution of the trained RC map at $\varepsilon = 3.856459$ closely mimicking the logistic map.}
    \label{fig8}
\end{figure}
To overcome this limitation, Ref.~\cite{sisodia2026finite} proposed the use of maximal FTLE distributions as an indirect diagnostic of the transition mechanism. The analysis focuses on the statistical distribution of FTLEs in the neighborhood of an interior crisis, which is characterized by a sudden enlargement of the chaotic attractor \cite{PhysRevLett.48.1507}. For the logistic map, this transition occurs at $\mu \approx 3.8568007$, where the chaotic attractor collides with an unstable period-3 orbit, resulting in an abrupt increase in attractor size. Near the crisis, the FTLE distribution (Fig.~\ref{fig8}a) consists of two statistically independent contributions \cite{PhysRevE.60.2761}: an exponential cusp corresponding to the pre-crisis attractor and a Gaussian component associated with trajectories exploring the enlarged chaotic attractor. A qualitatively identical behavior is observed in the trained reservoir map. At $\varepsilon = 3.856459$, the bifurcation diagram exhibits a sudden expansion of the chaotic attractor, indicating the occurrence of an interior crisis. The corresponding FTLE distribution of the trained reservoir (Fig.~\ref{fig8}b) closely reproduces that of the logistic map, including both the exponential cusp and Gaussian components. This strong agreement provides compelling indirect evidence that the learned reservoir undergoes an interior crisis through a dynamical mechanism analogous to that of the original system, despite the substantial difference in dimensionality between the two dynamical systems. Thus, the FTLE distribution serves as an effective dynamical signature of an interior crisis, eliminating the need for explicitly detecting periodic-orbit collisions. Having established this correspondence near the interior crisis, FTLE distributions are next examined over a wider range of bifurcation parameter values to determine whether the trained reservoir consistently reproduces the statistical signatures characteristic of different dynamical regimes.
\begin{figure}[t!]
    \centering
    \begin{tabular}{cc}
        \includegraphics[width=0.48\columnwidth]{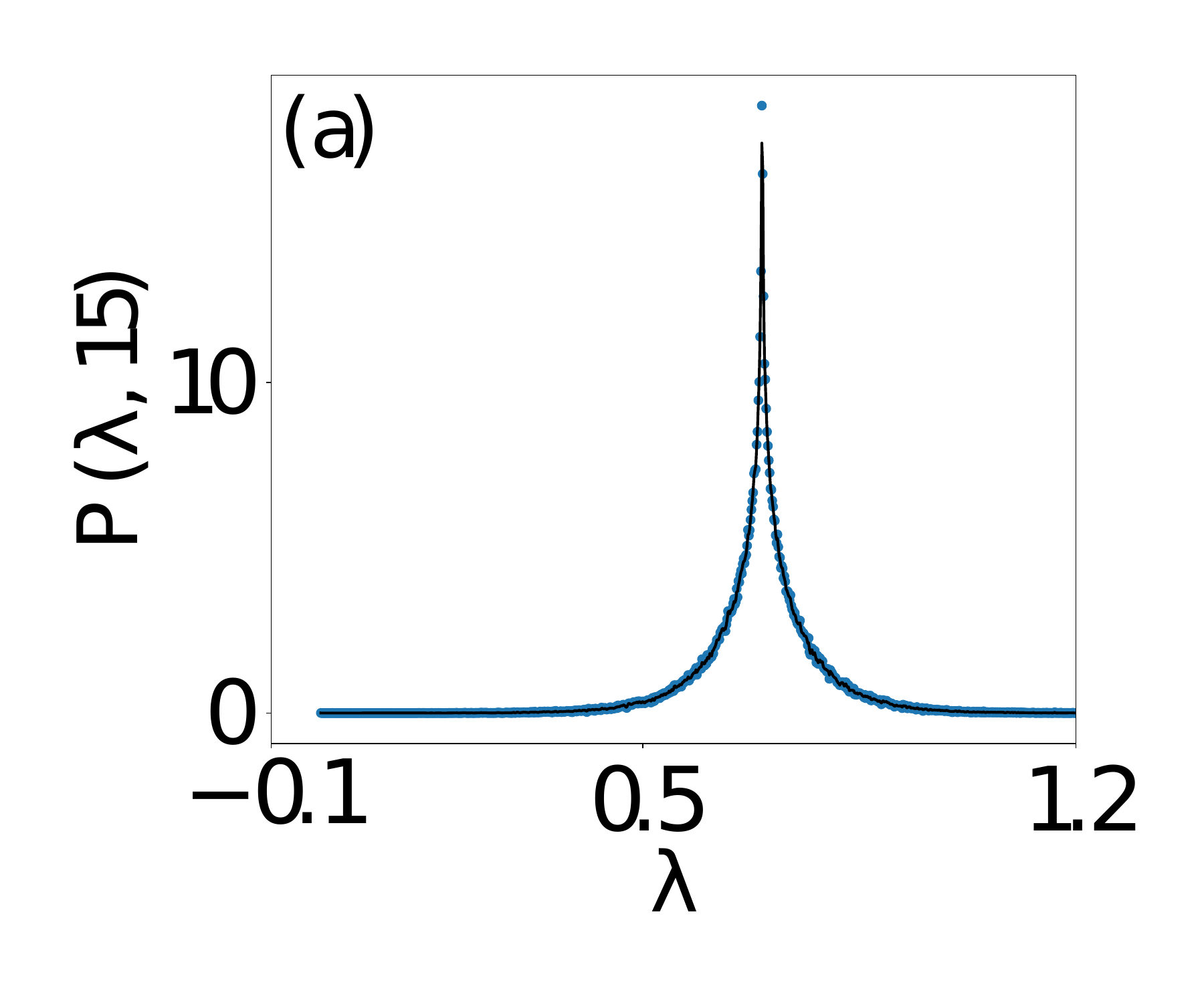} &
        \includegraphics[width=0.48\columnwidth]{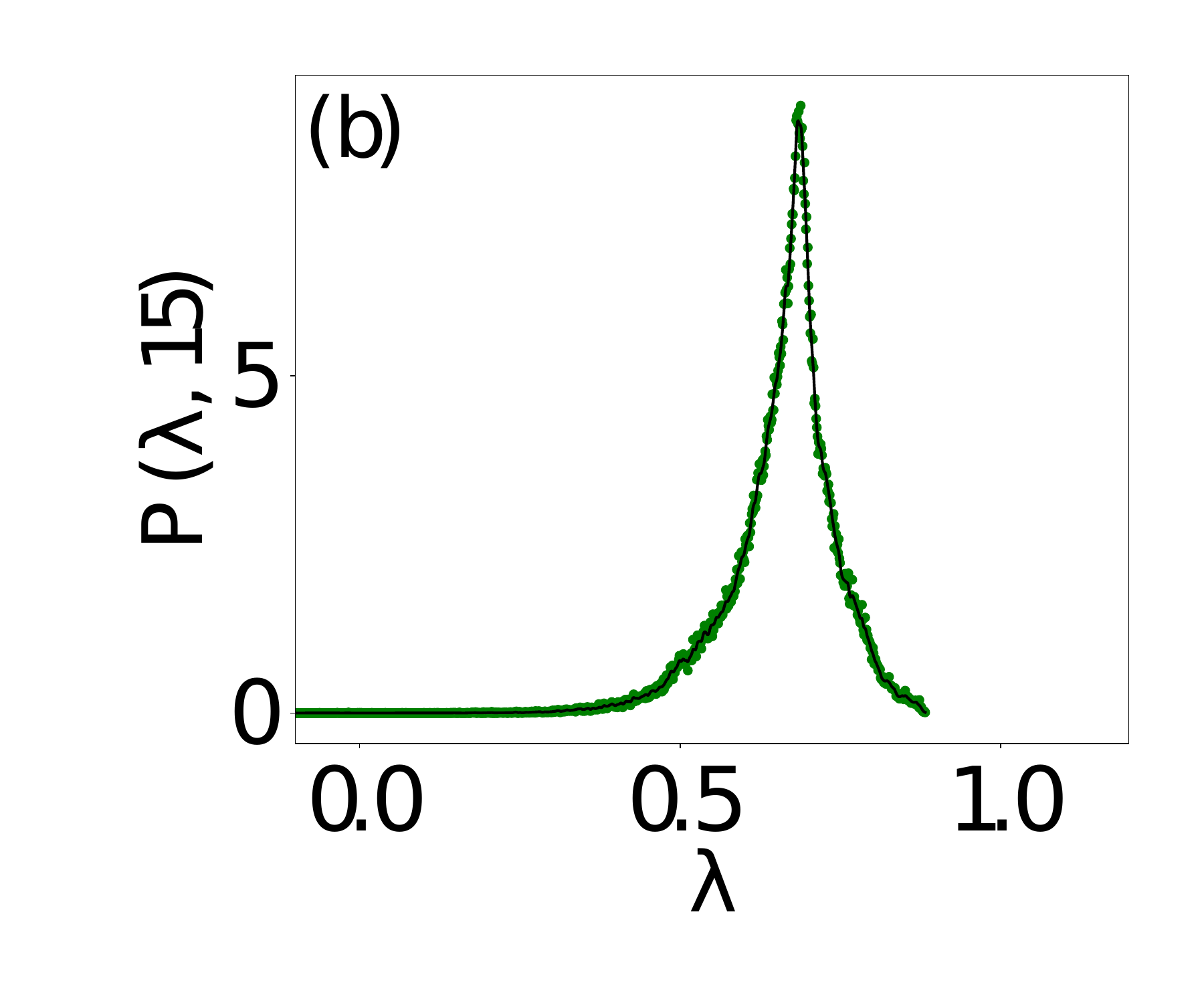} \\
        \includegraphics[width=0.48\columnwidth]{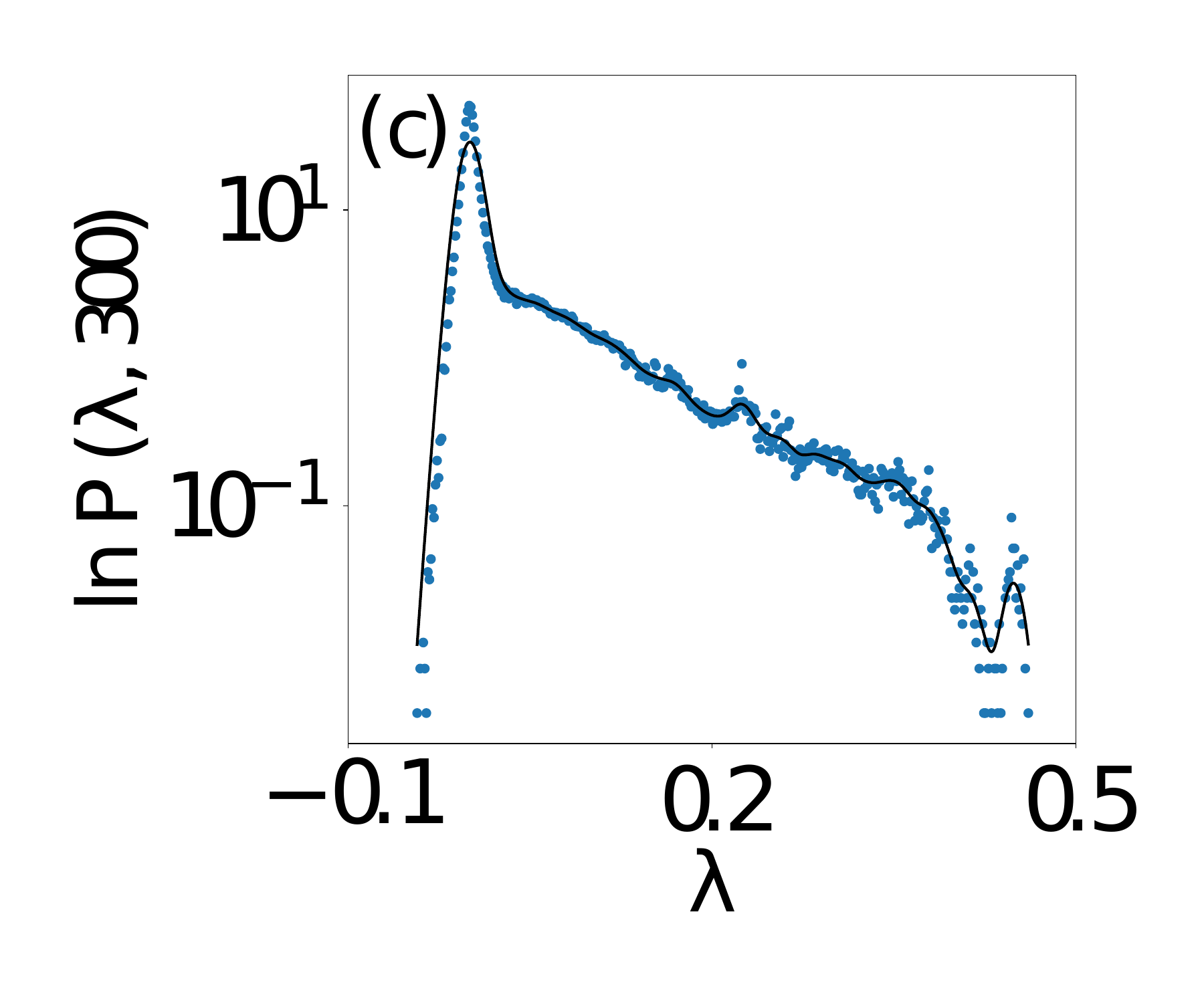} &
        \includegraphics[width=0.48\columnwidth]{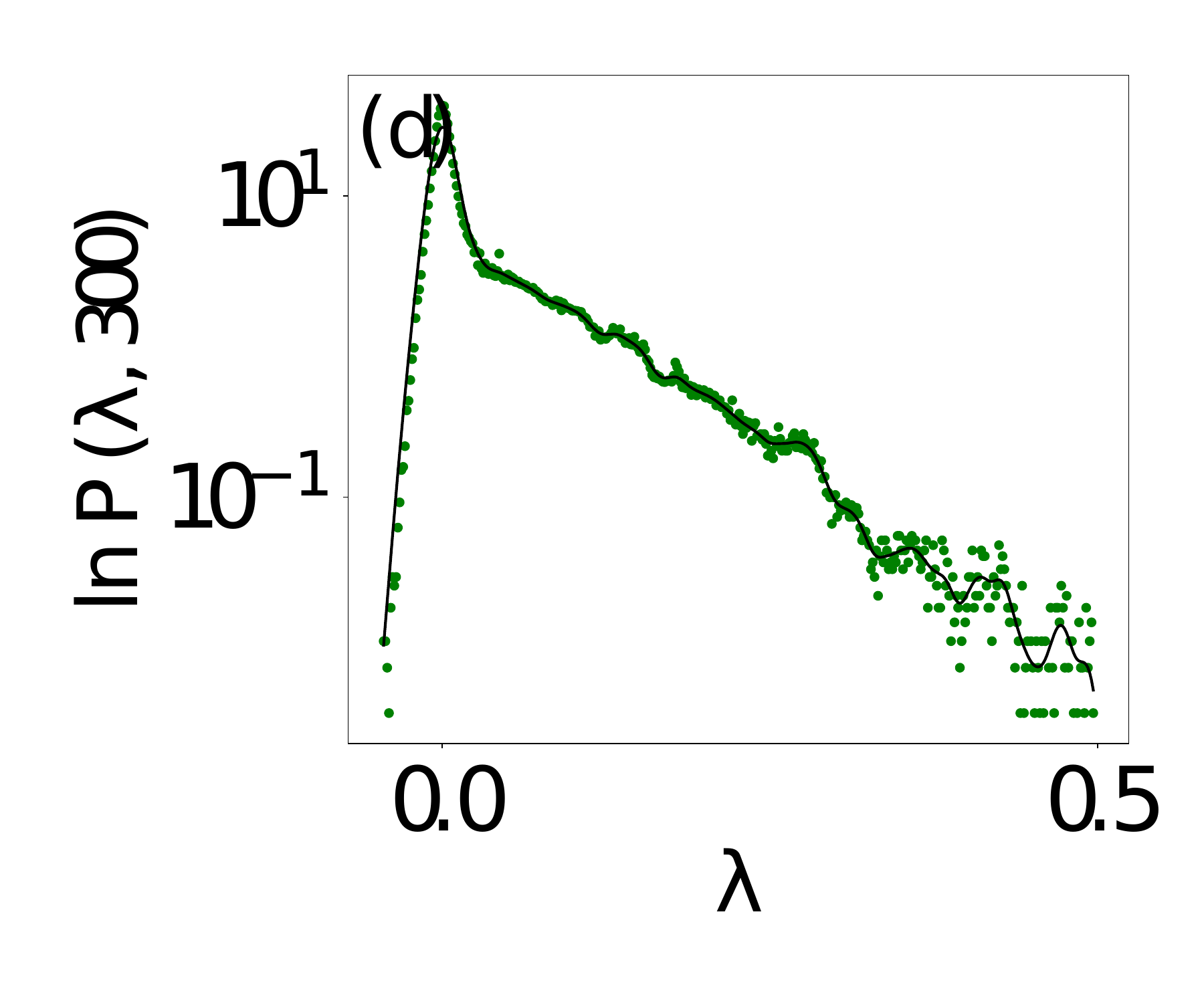}

    \end{tabular}
    \caption{Reproduced from \cite{sisodia2026finite}. (a) FTLE distribution for the logistic map at $\mu = 4$ (fully developed chaos) computed from trajectory segments of length $N = 15$. (b) Corresponding FTLE distribution for the trained RC map at $\varepsilon = 3.995$ using segments of length $N = 15$. The solid black curve represents the kernel density estimate of the underlying probability density. (c) FTLE distribution for the logistic map near the subduction at $\mu = 1 + \sqrt{8} - 10^{-4}$, computed using trajectory segments of length $N = 300$. (d) Corresponding FTLE distribution for the trained RC map at $\varepsilon = 3.827443$, computed using segments of length $N = 300$.}

    \label{fig9}
\end{figure}
Within the regime of fully developed chaotic dynamics ($3.57 < \mu < 4$), the rapid decay of temporal correlations causes the Lyapunov exponents to behave as random variables. As a consequence, FTLEs satisfy the central limit theorem, resulting in Gaussian probability distributions \cite{Ott2002_9}. These distributions are centered around $\lambda = \Lambda$, where $\Lambda$ denotes the asymptotic Lyapunov exponent obtained in the limit $t \to \infty$. At $\mu = 3.7$, both the logistic map and the trained reservoir map display the expected Gaussian FTLE distribution. A qualitatively different behavior emerges at the Ulam point ($\mu = 4$), corresponding to the regime of ``fully developed chaos." Here, the FTLE distribution departs significantly from the Gaussian form and instead develops a cusp singularity at $\lambda = \Lambda$, with $\Lambda = \ln 2$ representing the asymptotic Lyapunov exponent (Fig.~\ref{fig9}(a) and (b)). The analytical origin of these singular features has been established in Refs.~\cite{PhysRevE.60.2761,PhysRevE.69.016207}.
Another important transition exhibited by the logistic map is subduction, defined as the replacement of a chaotic attractor by a non-chaotic attractor while preserving its basin of attraction \cite{PhysRevLett.48.1507}. In the logistic map, this transition occurs at $\mu = 1+\sqrt{8}$, where the chaotic attractor is replaced by a stable period-3 orbit. A characteristic dynamical hallmark of subduction is the onset of intermittency in the time series \cite{RevModPhys.53.643}. In this regime, the FTLE probability distribution is characterized by a Gaussian core accompanied by a stretched-exponential tail (Fig.~\ref{fig9}c) \cite{PhysRevE.60.2761, Laiftle}. At the corresponding parameter value, $\varepsilon \approx 3.82744$, the trained reservoir map successfully reproduces not only the stretched-exponential tail (Fig.~\ref{fig9}d) but also its associated scaling exponent with good agreement.
\begin{figure}[t!]
     \includegraphics[width=\linewidth]{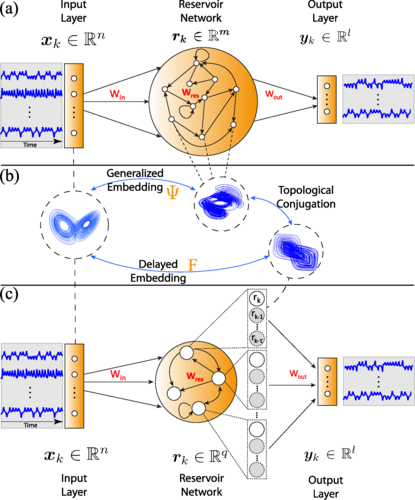}
    \caption{Reused from \cite{PhysRevResearch.5.L022041}. (a) A standard RC without time delay. (b) The generalized embedding $\Psi$ from the input dynamics to the standard non-delayed reservoir network and the delayed embedding $F$ from the input dynamics to the delayed reservoir network, which constitute a topological conjugation between the dynamics of the non-delayed reservoir network and the delayed reservoir network. (c) A time-delayed RC with a smaller network size in the reservoir layer.}
    \label{fig10}
\end{figure}
\subsection{Embedding theorems for reservoir}
Let the input sequence $u_0, u_1, u_2, \dots, u_k$ consist of one-dimensional observations originating from an invertible, discrete-time dynamical system. This system is governed by a continuously differentiable evolution operator $\phi \in \text{Diff}^1(M)$ acting on a compact, $n$-dimensional manifold $M$, where the system state is observed via a differentiable function $\omega \in C^1(M, \mathbb{R})$. 
An Echo State Network (ESN) is characterized by the triplet $(\varphi, A, W_{in})$, where $A \in \mathbb{R}^{m \times m}$ represents the reservoir adjacency matrix, $W_{in} \in \mathbb{R}^{m \times 1}$ is the input weight matrix, and $\varphi$ is a non-linear activation function (typically $\tanh$). The state update equation of the reservoir is expressed as:
\begin{equation}\label{eq5}
    r_{k+1} = \varphi(A r_k + W_{in} u_k)
\end{equation}
To formally analyze the state history, a family of functions $\mathcal{F} = \{f_k^{r_0} : M \to I^m \mid r_0 \in I^m, k \in \mathbb{N}_0\}$ is recursively defined by:
\begin{align}
    f_0^{r_0}(x) &= r_0 \nonumber \\
    f_{k+1}^{r_0}(x) &= \varphi\left(A f_k^{r_0}(\phi^{-1}(x)) + W_{in}\omega(x)\right) \nonumber
\end{align}
where $r_0$ represents the initial reservoir state. This formulation is mathematically equivalent to the state update rule (Eq.~\ref{eq5}). The function $f_k^{r_0}(x)$ evaluates the reservoir state after driving the ESN from an initial state $r_0$ using $k+1$ sequential observations: $\omega(\phi^{-k}(x)), \dots, \omega(\phi^{-1}(x)), \omega(x)$. Consequently, $f_k^{r_0}$ maps a state $x \in M$ to the reservoir space $I^m$ based entirely on its historical trajectory. 
Hart et al.~\cite{HART2020234} demonstrated that as the history length approaches infinity ($k \to \infty$), the dependency on the initial state vanishes for any arbitrary $r_0 \in I^m$. The limiting function:
\begin{equation}
    f = \lim_{k \to \infty} f_k^{r_0} \nonumber
\end{equation}
exists, is independent of $r_0$, and is formally defined as the \textit{Echo State Map}. The authors leverage this construction to prove the ``Weak ESN embedding theorem''~\cite{HART2020234}. The theorem establishes that if the reservoir dimension satisfies $m > 2n+1$, then for a generic observation function $\omega \in C^2(M, \mathbb{R})$, a randomly initialized Echo State Map $f$ is a $C^1$ embedding with a probability $\alpha > 0$, provided the following constraints hold:
\begin{itemize}
    \item  $\|A\|_2 < \min\left(\frac{1}{\|D\phi^{-1}\|_\infty}, 1\right)$
    where $\|A\|_2$ signifies the matrix 2-norm, and $\|D\phi^{-1}\|_\infty$ denotes the supremum norm of the Jacobian.
    
    \item The map $\phi \in \text{Diff}^2(M)$ possesses only finitely many periodic points with periods less than or equal to $2n$.
    
    \item For any periodic point $x \in M$ with period $k \leq 2n$, the eigenvalues of the derivative operator $D\phi^k$ evaluated at $x$ are distinct.
\end{itemize}
Once trained, the reservoir can be operated as an autonomous system by closing the loop with the learned output weights:
\begin{equation}
    r_{k+1} = \varphi\left((A + W_{in} W_{out})r_k\right) \nonumber
\end{equation}
The authors further prove that there exists a linear readout matrix $W_{out}$ that yields an autonomous ESN with dynamics that are topologically conjugate to those of the given dynamical system. 
A related theoretical development was presented in \cite{PhysRevResearch.5.L022041}. Let $\phi(\textbf{x}_t) = \textbf{x}_{t+1}$ denote the evolution operator of a discrete-time dynamical system, where $\textbf{x}_k \in \mathbb{R}^n$ is the system state evolving on a compact manifold $\mathcal{M}$, and $\phi\in\mathrm{Diff}^2(\mathcal{M})$. The reservoir update equation (Eq.~\ref{eq5}) can then be expressed as
\begin{equation}
    r_{k+1 , \textbf{x}_0} = \varphi(A r_{k, \textbf{x}_0} + W_{in}\phi^{k+1} \textbf{x}_0), \nonumber
\end{equation}
where the mapping $\mathcal{O}^k(r_0, A, W_{in})(\textbf{x}_0)$ is defined as $r_{k , \textbf{x}_0}$. It was shown that if the reservoir dimension satisfies $m \geq 2\mathrm{dim}(\mathcal{M})+1$, where $\mathrm{dim}(\mathcal{M})$ denotes the box-counting dimension of the manifold, then there exists an integer $k^*>0$ such that, for all $k>k^*$, the mapping $\mathcal{O}^k \in C^1(\mathcal{M},\mathbb{R}^m)$ is generically an embedding. Although in conventional RC, the dimension $m$ of the reservoir is typically large, the same work demonstrated that the required dimensionality can be substantially reduced by incorporating time delays in the output layer. Using delay embedding theory, it was proven that the resulting time-delayed reservoir-state representation also constitutes an embedding of the underlying dynamical system (Fig. \ref{fig10}), thereby providing a theoretically justified low-dimensional alternative to standard reservoir computing.

\subsection{Edge of Chaos, attractor reconstruction, and aDFA rule}
In the preceding sections, we have provided a detailed review of selected aspects of the functional understanding of RC from the perspective of nonlinear dynamics. In this section, we briefly address several additional studies that were not covered in the discussion above.

\paragraph{(i) Edge of chaos:} The idea that computation is richest near an order-disorder transition goes back to Langton's ``computation at the edge of chaos''~\cite{langton1990edge}, and was made concrete for recurrent networks by reservoir models; the liquid state machine~\cite{maass2002realtime} and the echo-state network~\cite{JaegerHaas2004}. The widely accepted view that reservoir computing (RC) achieves optimal computational performance exclusively at the edge of chaos was critically examined by Carroll~\cite{Carroll2020}. The study showed that this hypothesis is not universally valid, demonstrating that optimal performance can occur away from the edge of stability and identified the underlying mechanisms driving this behavior. Although the reservoir's entropy reaches its maximum near this boundary, it does not necessarily correspond to the highest prediction accuracy, highlighting that maximizing entropy alone is insufficient to optimize reservoir performance.

\paragraph{(ii) Attractor reconstruction and generation:}
Closely related to the theory of generalized synchronization \cite{PhysRevLett.76.1816}, it has been shown theoretically that a reservoir can embed the dynamics of a target system within its own state space \cite{lu2018attractor,grigoryeva2023learning}. More recently, numerous approaches have been proposed that exploit this capability to embed bifurcation structures, construct multi-attractors, or even generate novel bifurcations within reservoir systems. For example, Kim et al. demonstrated that, for the Lorenz system, providing few trajectories prior to a bifurcation enables the reservoir to extrapolate the entire bifurcation structure beyond the training regime \cite{kim2021teaching}. Following the concept of bifurcation extrapolation, Shen et al. showed that an ESN can successfully extrapolate the bifurcation structure of a passive dynamic walker \cite{shen2026predicting}. Akashi et al. showed that, in a physical reservoir based on a pneumatic artificial muscle, the external load can be treated as a bifurcation parameter. After embedding a single attractor, inducing a bifurcation enables the generation of a wide variety of attractors that were never presented during training \cite{akashi2024embedding}. Terajima et al., following the scheme of multifunctional reservoir computing \cite{flynn2021multifunctionality,flynn2023seeing}, demonstrated that a physical reservoir implemented on a tensegrity-structure simulator can embed multiple locomotion patterns, each represented as an autonomous dynamical system, using a single readout layer \cite{terajima2025multifunctional}. Kabayama et al. exploited the crisis bifurcation intrinsic to echo state networks (ESNs) and showed that, after embedding desired attractors, gradually increasing the spectral radius induces autonomous intermittent transitions among those attractors \cite{kabayama2025crisis,kabayama2025pseudo}. Akashi et al. showed that, noise-induced bifurcations (both noise-induced order and chaos) can be also extrapolated using ESN in a one-shot learning manner \cite{akashi2026one}. 
Terasaki et al. further focused on a one-dimensional reservoir and demonstrated that, after training only on a period three orbit, periodic orbits of all period, including chaotic dynamics, can be generated simply by varying the readout scaling after training so as to induce bifurcations \cite{terasaki2025thermodynamic}. Arun et al. demonstrated prediction of time series variables by using  Spin-torque nano oscillators as a reservoir and show its robustness in presence of thermal noise \cite{f2f3-l3fc}.

\paragraph{(iii) Generating chaotic attractor from unsuccessful learning:}
From the viewpoint of dynamical systems, unsuccessful learning is itself a consequence of the underlying dynamics rather than merely an optimization failure. In essence, the learning process simply modifies the system parameters through the interaction between the intrinsic properties of the untrained reservoir and the training data. 
Andrew et al. introduced the concept of untrained attractors to describe attractors that emerge when learning fails or latently exist, thereby broadening the conventional machine-learning perspective on what constitutes learning \cite{flynn2021multifunctionality,flynn2023seeing}. 
Building on this idea, Kabayama et al. showed that reservoirs operating near the boundary where the echo state property is marginally violated can intentionally destabilize a desired output template (referred to as a skeleton) and transform it into a chaotic attractor \cite{kabayama2025designing}. 
Remarkably, this approach can generate chaotic attractors even from hand-drawn geometric shapes for which no governing equations are available. 
In this sense, intentionally allowing learning to fail provides a creative use of untrained attractors, where chaos emerges not despite the failure of learning but precisely because of it.

\paragraph{(iv) Learning of physical systems through aDFA rule:}
Numerous gradient based optimization methods have been proposed for information processing devices involving physical systems \cite{wright2022deep,imai2025gradient,tomioka2025backpropagation}. 
However, such approaches inevitably face the sim-to-real gap, namely the discrepancy between the physical system and its mathematical model. 
Moreover, obtaining an accurate model of a physical reservoir; let alone its gradient information; is generally an extremely challenging task.
To address this issue, augmented direct feedback alignment (aDFA), a learning rule for deep physical reservoir computing that does not require gradient information, was proposed \cite{nakajima2022physical}. 
Instead of employing conventional error backpropagation, the transpose of the weight matrix appearing in backpropagation is replaced with a fixed random matrix \cite{lillicrap2016random}, allowing the output error to be propagated directly to each layer \cite{nokland2016direct}. 
Furthermore, the gradient information itself is replaced by an arbitrary nonlinear function \cite{nakajima2022physical}. 
These modifications make the learning algorithm highly suitable for implementation on physical hardware, where gradient computation is often infeasible.
Since its introduction, various extensions and variants of this framework have been proposed. For example, Zhang et al. systematically investigated the applicability of aDFA to deep spiking neural networks \cite{zhang2024training}. 
Nakajima et al. proposed a variant in which the backward matrix in aDFA is itself implemented as an ESN, aiming to further improve learning efficiency \cite{nakajima2024reservoir}. Furthermore, Zhang et al. introduced a state-skipping approach, thereby extending both the applicability and effectiveness of aDFA in physical deep learning \cite{zhang2026scalable}.

\section{A brief survey on LSTM models as dynamical systems}
\label{sec:lstm}
The preceding sections focused on reservoir computing, where the recurrent weights remain fixed after initialization and only the output layer is trained. However, the dynamical systems framework is equally applicable to RNNs with trainable recurrent connections. Unlike RC, the underlying dynamics of these networks evolve continuously during optimization as the recurrent weights are updated through gradient descent. Consequently, the network traverses different dynamical regimes throughout training, allowing its computational performance to be examined in relation to the corresponding phase diagram. This perspective has been explored for LSTM networks \cite{hochreiter1997lstm}. 
\begin{figure*}[ht!]
  \centering
  \includegraphics[width=0.85\textwidth]{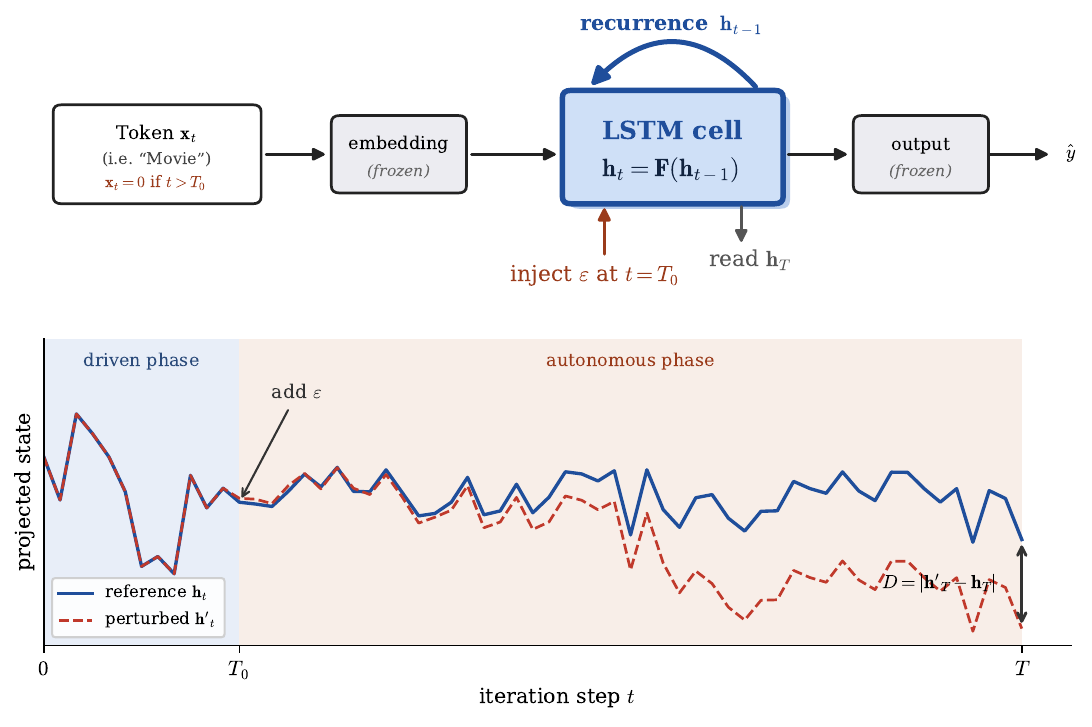}
  \caption{Probing the intrinsic dynamics of a trained LSTM cell (method use in  Ref.~\cite{wei2026multiple}).
  \emph{Top:} the measurement pipeline. The input is the token $\bm{x}_t$
  during the driven phase ($t\le T_0$) and the zero vector thereafter
  ($t>T_0$, autonomous phase); it passes through a frozen (not trained) embedding layer to
  the cell. The recurrent LSTM cell, $\bm{h}_t=\bm{F}(\bm{h}_{t-1})$, is the
  only evolving map (emphasized), while the surrounding embedding and output
  (readout) layers are held fixed. A small perturbation $\bm{\varepsilon}$ is
  injected into the hidden state $\bm{h}_{t}$ at $t=T_0$, and the asymptotic hidden state
  $\bm{h}_{T}$ is read out at a large step $T$. \emph{Bottom:} the reference
  and perturbed hidden trajectories (here shown for a chaotic case) coincide
  through the driven phase and separate once the input is removed; their
  final separation defines the asymptotic distance
  $D=|\bm{h}_T'-\bm{h}_T|$, which diagnoses whether the trained cell is in
  the ordered or chaotic phase.}
  \label{fig:lstm-schematic}
\end{figure*}

\subsection{The LSTM cell as a recurrent map}
\label{sec:lstm-map}
To allow recurrent networks retain information over long time lags \cite{hochreiter1997lstm}, an LSTM cell propagates two state vectors; a hidden state $\bm{h}_t$ and a cell (memory) state $\bm{c}_t$; which are updated from their previous values along with the current input $\bm{x}_t$ (which can be a word in a sentence, or stock price at a given time)
through the gated, mostly hyperbolic nonlinearities~\cite{hochreiter1997lstm}.
\begin{align}
  \bm{i}_t        &= \sigma_s\!\left(\bm{W}_i \bm{x}_t + \bm{U}_i \bm{h}_{t-1} + \bm{b}_i\right), \nonumber\\
  \bm{f}_t        &= \sigma_s\!\left(\bm{W}_f \bm{x}_t + \bm{U}_f \bm{h}_{t-1} + \bm{b}_f\right), \nonumber\\
  \bm{o}_t        &= \sigma_s\!\left(\bm{W}_o \bm{x}_t + \bm{U}_o \bm{h}_{t-1} + \bm{b}_o\right), \nonumber\\
  \tilde{\bm{c}}_t &= \sigma_t\!\left(\bm{W}_c \bm{x}_t + \bm{U}_c \bm{h}_{t-1} + \bm{b}_c\right),
  \label{eq6}\\
  \bm{c}_t        &= \bm{f}_t \odot \bm{c}_{t-1} + \bm{i}_t \odot \tilde{\bm{c}}_t, \nonumber\\
  \bm{h}_t        &= \bm{o}_t \odot \sigma_t(\bm{c}_t), \nonumber
\end{align}
with $\sigma_s$, $\sigma_t$ being the sigmoid and the hyperbolic tangent activations respectively, $\odot$ the
element-wise product, and learned weights and biases $\bm{W},\bm{U},\bm{b}$.
The input, forget and output gates $\bm{i}_t,\bm{f}_t,\bm{o}_t$ do not act as independent variables but as a state-dependent, structured nonlinearity which regulates how the memory is written, retained and exposed. The update equations in Eqs.~\ref{eq6} can be viewed as a high-dimensional generalization of the recurrence in Eq.~\ref{eq2}. By identifying $\bm{k}_t \equiv \bm{h}_t$, the entire LSTM cell may be regarded as a single nonlinear map $\bm{F}$. (The full internal state of an LSTM is the pair $  (\bm{h_t}, \bm{c_t})  $; here we treat the hidden state as the primary dynamical variable, consistent with common practice when analyzing the intrinsic dynamics of trained networks.) Although this map is considerably more intricate than the RC map discussed earlier, it can be analyzed using the dynamical systems tools.

A practical consideration arises when applying these tools to LSTM models. Unlike autonomous dynamical systems, recurrent networks are typically \emph{driven} by an external input sequence $\bm{x}_t$, whereas the order or chaos of a map is an \emph{intrinsic} property to be extracted from $\bm{h}_t$ defined in the absence of external forcing. To disentangle these two effects \cite{wei2026multiple}, the network dynamics can be divided into a driven phase and an autonomous phase (Fig.~\ref{fig:lstm-schematic}). During the driven phase, the LSTM processes the input sequence (e.g., the embedded tokens of a movie review in a sentiment classification task) for the first $T_0$ time steps, allowing the trained dynamics to converge to their typical operating region. At time $T_0$, a small perturbation is applied to the hidden state $\bm{h}_{T_0}$, after which the external input is removed by setting $\bm{x}_{t>T_0}=\bm{0}$. The network is then allowed to evolve autonomously until a sufficiently large final time $T$. With the input stream disabled and the embedding and output layers kept fixed, the recurrent LSTM block evolves solely according to $\bm{h}_t=\bm{F}(\bm{h}_{t-1})$, so its long-term dynamics reveal the intrinsic phase of the recurrent cell. This autonomous relaxation is directly analogous to the regime reached by a reservoir once transient input has decayed, enabling trained LSTMs and fixed reservoirs to be analyzed within the same dynamical framework.

\paragraph{Reading the phase:}
\label{sec:lstm-phase}
With the recurrent dynamics isolated in this way, the diagnostics introduced in Sec.~\ref{sec:tools} can be applied directly. The most straightforward is the asymptotic distance described in Sec.~\ref{sec:asymp-distance}. A small perturbation is applied to the hidden state at the beginning of the autonomous phase, $\bm{h}_{T_0}'=\bm{h}_{T_0}+\bm{\varepsilon}$, and both the perturbed and unperturbed trajectories are evolved until a sufficiently large time $T$. Their final separation, $D = |\bm{h}_T' -
\bm{h}_T|$ and its ensemble geometric-mean logarithm, $\widetilde{D}$, provide a direct indicator of the order-chaos state of the model throughout training, within the machine-precision baseline. Since $\widetilde{D}$ indicates only whether nearby trajectories ultimately diverge, it is naturally complemented by a measure of the rate of divergence, provided by the finite-time Lyapunov exponents (FTLEs) introduced in Sec.~\ref{sec:tools}, whose sign distinguishes the two dynamical phases: negative values correspond to ordered dynamics, whereas positive values indicate chaos. The transition between the two regimes is therefore identified by the FTLE crossing zero. Together, the asymptotic distance and the FTLE offer complementary information: the former provides a computationally efficient and robust means of locating the transition, while the latter quantifies the local dynamical instability in its vicinity. A schematic of the perturbation procedure is shown in Fig.~\ref{fig:lstm-schematic}.
Extracting the bifurcation structure can also be visually clear at discerning the phases, but requires one further step. The hidden
state is high-dimensional, so typical bifurcations cannot be displayed in the usual one-dimensional bifurcation diagram. One can therefore project the asymptotic state onto a single scalar by the reduced sum ~\cite{wei2026multiple}:
\begin{equation}
  s \;=\; \bm{h}_T \cdot \bm{1} \;=\; \sum_{j} (\bm{h}_T)_j , \nonumber
  \label{eq:reduced-sum}
\end{equation}
i.e.\ the sum of the components of the asymptotic hidden vector. Plotting
$s$ across an ensemble of inputs against a control parameter recovers a
one-dimensional bifurcation diagram for the LSTM: in the ordered phase the
samples collapse onto a single value, while in the chaotic phase they
spread out over a finite range. This reduction of a high-dimensional state to a single informative coordinate is motivated by the same principle used when projecting high-dimensional reservoir states. It is this dimensionality reduction that makes the bifurcation structure of a network with tens of hidden units amenable to visualization and interpretation.

\begin{figure}[t]
  \centering
  \includegraphics[width=\linewidth]{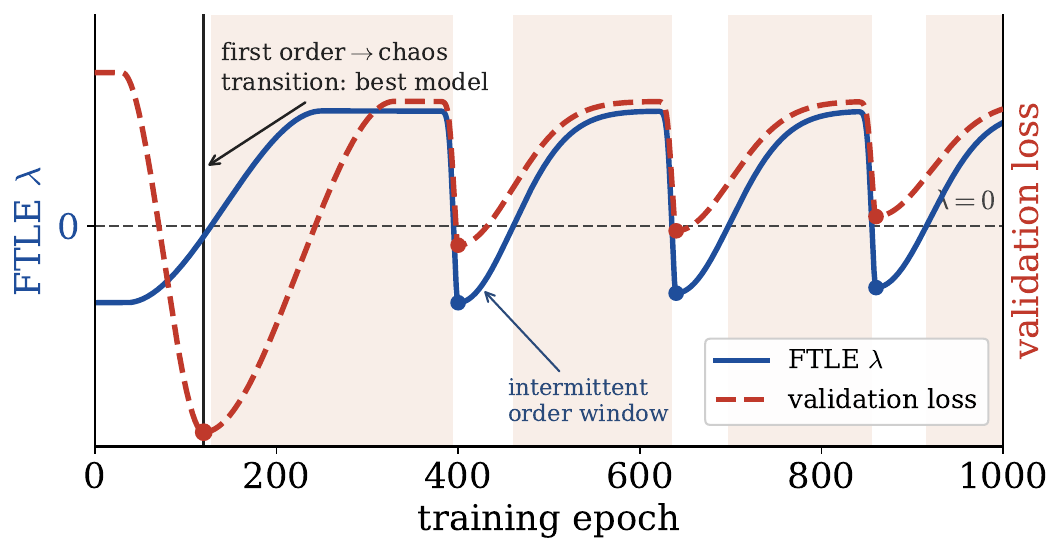}
  \caption{Qualitative relationship between the order and chaos states of a
  LSTM and its validation loss over training (schematic representation of the results from
  Ref.~\cite{wei2026multiple}). The finite-time Lyapunov exponent (FTLE
  $\lambda$, blue) starts negative (ordered) and rises gradually through the
  boundary $\lambda=0$ at the \emph{first}, order-to-chaos transition; the
  validation loss (red, dashed) descends gradually to its \emph{global}
  minimum there (\emph{global} as for that training realization), marking the best-performing epoch. In the subsequent
  overtrained regime the dynamics are mostly chaotic (shaded), punctuated by
  brief intermittent windows of order: at each stability transition both curves can drop sharply to a
  coincident local minimum and then recover gradually.}
  \label{fig:eoc-loss}
\end{figure}

\paragraph{Edge of chaos and predictive efficiency:}
\label{sec:lstm-eoc}
Tracking $\widetilde{D}$, the FTLE, and the projected bifurcation diagram throughout training reveals that an LSTM does not remain confined to a single dynamical phase. In the absence of explicit regularization, stochastic gradient descent gradually increases the weight magnitudes, causing the network to cross the order/chaos transition boundary multiple times and undergo a sequence of stability transitions~\cite{wei2026multiple}. Stability analysis further shows that the epochs with the locally optimal generalization generally coincide with these transitions (Fig.~\ref{fig:eoc-loss}). In particular, the validation loss attains its best minimum at the \emph{first} order-to-chaos transition, where the edge is widest and the exploration of weight configurations is greatest, while additional local minima appear at subsequent transitions~\cite{wei2026multiple}. From a dynamical-systems perspective, this is the observed ``multiple-descent'' behavior of the loss curve. An important note is that, the descent in loss function needs not to be only associated with order to chaos transitions or vice versa, but can be associated with transitions among different chaotic attractors as well.

\subsection{From recurrent to feed-forward networks}
\label{sec:lstm-ff}
The dynamical systems perspective is not confined to recurrent architectures. A feed-forward network can also be viewed as an iterable map by selecting a hidden layer whose input and output have the same dimension (typically the layer immediately preceding the final softmax). The network up to this layer then defines a nonlinear operator $\bm{F}$, whose asymptotic stability can be analyzed using the same tools described above \cite{feng2020edge}. This approach reveals the same three dynamical regimes: fixed points, periodic orbits, and chaos and shows that well-performing computer vision models also operate near the edge of chaos, with training naturally driving the dynamics toward this regime. In certain cases, the resulting order-chaos phase diagram can even be related to that of the Sherrington-Kirkpatrick spin-glass model, providing a dynamical interpretation of the edge-of-chaos criterion and offering practical guidance for selecting hyperparameters such as weight decay \cite{zhang2024asymptotic}. Recurrent and feed-forward networks therefore fall within a common dynamical-systems framework, although RC and RNNs remain its most natural representatives because their dynamics arise from explicit iteration.

\section{Code Repository and Implementation Details}
To bridge the theoretical frameworks presented throughout this review with hands-on numerical experimentation, an open-source GitHub repository accompanies this article \cite{reservoir_computing_tutorial_code}. The repository serves as an accessible entry point for readers seeking to understand RC from a dynamical systems perspective. Designed with a modular architecture, each notebook is entirely self-contained and can be executed independently without requiring prior file executions. Furthermore, comprehensive markdown cells detailing the underlying equations and theoretical context precede the cells, wherever required, enabling readers to seamlessly reproduce and extend the numerical analyses described in the preceding sections.

\subsection{Organization of the Supplementary Notebooks}
The code is structured into eight standalone Jupyter notebooks. These notebooks are divided into two parts: fundamental nonlinear dynamics diagnostic tools (Notebooks 01-05) and their direct application to probing trained RC architectures (Notebooks 06-08).

\subsubsection{Foundational Dynamics Diagnostics (Notebooks 01-05)}
\begin{itemize}
    \item \textbf{\texttt{01\_Jacobian\_eigenvalues.ipynb}:} Demonstrates local stability analysis and bifurcation tracking by monitoring the trajectory of Jacobian eigenvalues across the unit circle. Taking a two-dimensional map as an example, it illustrates a Neimark-Sacker bifurcation where a pair of complex conjugate eigenvalues crosses the unit circle, resulting in the loss of stability of a fixed point and the birth of an invariant closed curve (supplementing the theory in Sec. \ref{subsec:eig}).
    \item \textbf{\texttt{02\_Bifurcations\_crisis.ipynb}:} Provides generalized scripts to compute and plot bifurcation diagrams for both continuous-time (Lorenz attractor) and discrete-time (Logistic map) dynamical systems (described in Sec. \ref{subsec:bif}).
    \item \textbf{\texttt{03\_FTLE.ipynb}:} Implements Finite-Time Lyapunov Exponents (FTLEs) to quantify local stability variations over finite time horizons, displaying FTLE distribution profiles across different dynamical regimes of a logistic map (described in Sec. \ref{subsec:ftle}).
    \item \textbf{\texttt{04\_scaling\_transient\_times.ipynb}:} Investigates boundary crises where trajectories display chaotic transients near the remnant of a destroyed attractor. The notebook numerically verifies the characteristic power-law scaling of average transient lifetimes prior to escape (described in Sec. \ref{subsec:scaling}).
    \item \textbf{\texttt{05\_delay\_coordinate\_embedding.ipynb}:} Implements Takens' delay coordinate embedding technique using a single scalar output variable from the Lorenz system. It demonstrates that the topology and geometry of a multi-dimensional chaotic attractor can be faithfully reconstructed from one-dimensional observations (described in Sec. \ref{subsec:delay}).
\end{itemize}

\subsubsection{Reservoir Computing Diagnostics and Analysis (Notebooks 06--08)}
\begin{itemize}
    \item \textbf{\texttt{06\_RC\_crisis.ipynb}:} Trains an RC model on the Logistic map for prediction of crisis and probes the dynamics of trained RC by a projection to low dimensions (supplementing Sec. \ref{subsec:RCCrisis}).
    \item \textbf{\texttt{07\_RC\_FTLE.ipynb}:} Evaluates the FTLE distributions of the autonomous map generated by an RC trained on the Logistic map, providing a complementary analysis of RC dynamics (supplementing Sec. \ref{subsec:RCFtle}).
    \item \textbf{\texttt{08\_RC\_Jacobian\_eigvals.ipynb}:} Trains an RC model on the Stuart-Landau oscillator for prediction of amplitude death and computes the Jacobian eigenvalues of the trained RC to track the dynamical transitions inside the reservoir (supplementing Sec. \ref{subsec:RCjac}).
\end{itemize}

\section{Conclusion}
The article offers an in-depth overview of recent progress in understanding how RC operates from the standpoint of dynamical systems theory. We introduced and implemented a broad suite of nonlinear dynamical analysis methods, along with the associated codes, and used them to characterize the behavior and functioning of RC. In particular, we showed that RC can learn and foresee critical transitions in an input system by internalizing its underlying dynamical and statistical structure, thereby forming a representation that goes beyond simple prediction. A central point we highlight is that, by treating the iterative update rule of RC as a high dimensional map, one can employ the full arsenal of statistical and dynamical tools developed for standard nonlinear dynamical systems to gain a deeper understanding of the internal representations encoded by the RC.

\backmatter

\bmhead{Acknowledgments}
S.J. gratefully acknowledges SERB Power grant award SPF/2021/000136. D.S. thanks CSIR fellowship under award no. JRF-NET2024/14347.
K.N. was supported by JSPS KAKENHI (Grant No. 25H01134) and by a project, JPNP14004, commissioned by the New Energy and Industrial Technology Development Organization
(NEDO). 

\bmhead{Declarations}

\begin{itemize}

\item Conflict of interest: The authors declare no competing interest.
\item Code availability statement: The supplementary Jupyter notebooks are openly available on GitHub (\url{https://github.com/complex-systems-lab/Tutorial_Review_Reservoir_Computing})

\end{itemize}

\noindent

\bibliography{references}

@PREAMBLE{
"\providecommand{\noopsort}[1]{}" 
# "\providecommand{\singleletter}[1]{#1}%" 
}

@article{nakajima2022physical,
  title={Physical deep learning with biologically inspired training method: gradient-free approach for physical hardware},
  author={Nakajima, Mitsumasa and Inoue, Katsuma and Tanaka, Kenji and Kuniyoshi, Yasuo and Hashimoto, Toshikazu and Nakajima, Kohei},
  journal={Nature Communications},
  volume={13},
  number={1},
  pages={7847},
  year={2022},
  publisher={Nature Publishing Group UK London}
}

@article{nakajima2024reservoir,
  title={Reservoir direct feedback alignment: Deep learning by physical dynamics},
  author={Nakajima, M. and Zhang, Y. and Inoue, K. and Kuniyoshi, Y. and Hashimoto, T. and Nakajima, K.},
  journal={Communications Physics},
  volume={7},
  number={1},
  pages={411},
  year={2024}
}

@article{zhang2026scalable,
  title={Scalable physical deep learning using optical dynamics with state-skipping training and single-map architecture},
  author={Zhang, Y. and Nakajima, M. and Inoue, K. and Hashimoto, T. and Kuniyoshi, Y. and Nakajima, K.},
  journal={Nature Computational Science},
  year={2026},
  note={accepted}
}

@article{terajima2025multifunctional,
  title={Multifunctional physical reservoir computing in soft tensegrity robots},
  author={Terajima, R. and Inoue, K. and Nakajima, K. and Kuniyoshi, Y.},
  journal={Chaos},
  volume={35},
  pages={083111},
  year={2025}
}

@article{akashi2024embedding,
  title={Embedding bifurcations into pneumatic artificial muscle},
  author={Akashi, N. and Kuniyoshi, Y. and Jo, T. and Nishida, M. and Sakurai, R. and Wakao, Y. and Nakajima, K.},
  journal={Advanced Science},
  volume={11},
  pages={2304402},
  year={2024}
}

@article{imai2025gradient,
  title={Gradient-based optimization of spintronic devices},
  author={Imai, Y. and Liu, S. and Akashi, N. and Nakajima, K.},
  journal={Applied Physics Letters},
  volume={126},
  number={8},
  year={2025}
}

@article{tomioka2025backpropagation,
  title={Backpropagation Through Soft Body: Investigating Information Processing in Brain--Body Coupling Systems},
  author={Tomioka, H. and Inoue, K. and Kuniyoshi, Y. and Nakajima, K.},
  journal={Advanced Robotics Research},
  pages={e202500159},
  year={2025}
}

@article{wright2022deep,
  title={Deep physical neural networks trained with backpropagation},
  author={Wright, Logan G and Onodera, Tatsuhiro and Stein, Martin M and Wang, Tianyu and Schachter, Darren T and Hu, Zoey and McMahon, Peter L},
  journal={Nature},
  volume={601},
  number={7894},
  pages={549--555},
  year={2022},
  publisher={Nature Publishing Group UK London}
}

@article{akashi2026one,
  title={One-shot prediction of noise-induced bifurcations with reservoir computing},
  author={Akashi, N. and Watanabe, T. and Hara, M. and Namiki, T. and Kokubu, H. and Tsuda, I. and Nakajima, K.},
  journal={arXiv preprint arXiv:2606.26727},
  year={2026}
}

@article{shen2026predicting,
  title={Predicting Bifurcation of Mechanical Systems Using Reservoir Computing: Case Studies on Legged Locomotion and Pneumatic Soft Actuator},
  author={Shen, J. and Al Ajroudi, R. and Akashi, N. and Jo, T. and Nishida, M. and Wakao, Y. and Sakurai, R. and Kuniyoshi, Y. and Nakajima, K.},
  journal={Chaos},
  volume={36},
  pages={013103},
  year={2026}
}

@article{kabayama2025crisis,
  title={Crisis-induced intermittency in reservoir computing},
  author={Kabayama, T. and Komuro, M. and Kuniyoshi, Y. and Aihara, K. and Nakajima, K.},
  journal={Physical Review Research},
  volume={7},
  pages={L032058},
  year={2025}
}

@article{kabayama2025pseudo,
  title={Pseudo-Lorenz Attractors in Echo State Networks},
  author={Kabayama, T. and Komuro, M. and Aihara, K. and Nakajima, K.},
  journal={Japan Journal of Industrial and Applied Mathematics},
  volume={42},
  pages={1459--1473},
  year={2025}
}

@article{terasaki2025thermodynamic,
  title={Thermodynamic limit in learning period three},
  author={Terasaki, Y. and Nakajima, K.},
  journal={Physical Review Research},
  volume={7},
  pages={023119},
  year={2025}
}

@article{kabayama2025designing,
  title={Designing chaotic attractors: A semisupervised approach},
  author={Kabayama, T. and Kuniyoshi, Y. and Aihara, K. and Nakajima, K.},
  journal={Physical Review E},
  volume={111},
  pages={034207},
  year={2025}
}

@article{lillicrap2016random,
  title={Random synaptic feedback weights support error backpropagation for deep learning},
  author={Lillicrap, T. P. and Cownden, D. and Tweed, D. B. and Akerman, C. J.},
  journal={Nature Communications},
  volume={7},
  number={1},
  pages={13276},
  year={2016}
}

@article{zhang2024training,
  title={Training spiking neural networks via augmented direct feedback alignment},
  author={Zhang, Y. and Inoue, K. and Nakajima, M. and Hashimoto, T. and Kuniyoshi, Y. and Nakajima, K.},
  journal={arXiv preprint arXiv:2409.07776},
  year={2024}
}

@inproceedings{nokland2016direct,
  title={Direct feedback alignment provides learning in deep neural networks},
  author={N{\o}kland, A.},
  booktitle={Advances in Neural Information Processing Systems},
  volume={29},
  year={2016}
}

@article{grigoryeva2023learning,
  title={Learning strange attractors with reservoir systems},
  author={Grigoryeva, L. and Hart, A. and Ortega, J. P.},
  journal={Nonlinearity},
  volume={36},
  number={9},
  pages={4674--4708},
  year={2023}
}

@article{kim2021teaching,
  title={Teaching recurrent neural networks to infer global temporal structure from local examples},
  author={Kim, J. Z. and Lu, Z. and Nozari, E. and Pappas, G. J. and Bassett, D. S.},
  journal={Nature Machine Intelligence},
  volume={3},
  number={4},
  pages={316--323},
  year={2021}
}

@article{flynn2021multifunctionality,
  title={Multifunctionality in a reservoir computer},
  author={Flynn, A. and Tsachouridis, V. A. and Amann, A.},
  journal={Chaos: An Interdisciplinary Journal of Nonlinear Science},
  volume={31},
  number={1},
  pages={013118},
  year={2021}
}

@article{flynn2023seeing,
  title={Seeing double with a multifunctional reservoir computer},
  author={Flynn, A. and Tsachouridis, V. A. and Amann, A.},
  journal={Chaos: An Interdisciplinary Journal of Nonlinear Science},
  volume={33},
  number={11},
  pages={113127},
  year={2023}
}

@book{nakajima2021reservoir,
  title={Reservoir Computing},
  author={Nakajima, Kohei and Fischer, Ingo},
  year={2021},
  publisher={Springer}
}

@article{lu2018attractor,
  title={Attractor reconstruction by machine learning},
  author={Lu, Zhixin and Hunt, Brian R and Ott, Edward},
  journal={Chaos: An Interdisciplinary Journal of Nonlinear Science},
  volume={28},
  number={6},
  pages={061104},
  year={2018},
  publisher={AIP Publishing},
  doi={10.1063/1.5039508}
}

@article{feng2020edge,
  author        = {Feng, Ling and Zhang, Lin and Lai, Choy Heng},
  title         = {Optimal Machine Intelligence at the Edge of Chaos},
  journal       = {arXiv preprint arXiv:1909.05176},
  year          = {2020},
  note          = {Submitted 2019; revised 2020 (v2)},
  eprint        = {1909.05176},
  archivePrefix = {arXiv},
  primaryClass  = {cs.LG},
  doi           = {10.48550/arXiv.1909.05176}
}

@article{zhang2024asymptotic,
  author  = {Zhang, Lin and Feng, Ling and Chen, Kan and Lai, Choy Heng},
  title   = {Asymptotic Edge of Chaos as Guiding Principle for Neural
             Network Training},
  journal = {International Journal of Artificial Intelligence and Robotics
             Research},
  volume  = {1},
  number  = {1},
  pages   = {2350001},
  year    = {2024},
  doi     = {10.1142/S2972335323500011}
}

@article{wei2026multiple,
  author  = {Wei, Wenbo and Xu, Fan and Chong, Nicholas Jia Le and
             Lai, Choy Heng and Feng, Ling},
  title   = {Multiple Descents in Deep Learning as a Sequence of
             Order--Chaos Transitions in {LSTM} Networks},
  journal = {Physica A: Statistical Mechanics and its Applications},
  volume  = {697},
  pages   = {131754},
  year    = {2026},
  doi     = {10.1016/j.physa.2026.131754}
}

@article{eckmann1985ergodic,
  author  = {Eckmann, Jean-Pierre and Ruelle, David},
  title   = {Ergodic Theory of Chaos and Strange Attractors},
  journal = {Reviews of Modern Physics},
  volume  = {57},
  number  = {3},
  pages   = {617--656},
  year    = {1985},
  doi     = {10.1103/RevModPhys.57.617}
}

@book{strogatz2015nonlinear,
  author    = {Strogatz, Steven H.},
  title     = {Nonlinear Dynamics and Chaos: With Applications to Physics,
               Biology, Chemistry, and Engineering},
  edition   = {2nd},
  publisher = {Westview Press},
  address   = {Boulder, CO},
  year      = {2015}
}

@article{langton1990edge,
  author  = {Langton, Christopher G.},
  title   = {Computation at the Edge of Chaos: Phase Transitions and
             Emergent Computation},
  journal = {Physica D: Nonlinear Phenomena},
  volume  = {42},
  number  = {1--3},
  pages   = {12--37},
  year    = {1990},
  doi     = {10.1016/0167-2789(90)90064-V}
}

@article{maass2002realtime,
  author  = {Maass, Wolfgang and Natschl{\"a}ger, Thomas and
             Markram, Henry},
  title   = {Real-Time Computing without Stable States: A New Framework
             for Neural Computation Based on Perturbations},
  journal = {Neural Computation},
  volume  = {14},
  number  = {11},
  pages   = {2531--2560},
  year    = {2002},
  doi     = {10.1162/089976602760407955}
}

@article{hochreiter1997lstm,
  author  = {Hochreiter, Sepp and Schmidhuber, J{\"u}rgen},
  title   = {Long Short-Term Memory},
  journal = {Neural Computation},
  volume  = {9},
  number  = {8},
  pages   = {1735--1780},
  year    = {1997},
  doi     = {10.1162/neco.1997.9.8.1735}
}

@article{10.1063/5.0016505,
    author = {Tang, Yang and Kurths, Jürgen and Lin, Wei and Ott, Edward and Kocarev, Ljupco},
    title = "{Introduction to Focus Issue: When machine learning meets complex systems: Networks, chaos, and nonlinear dynamics}",
    journal = {Chaos: An Interdisciplinary Journal of Nonlinear Science},
    volume = {30},
    number = {6},
    pages = {063151},
    year = {2020},
    month = {06},
    issn = {1054-1500},
    doi = {10.1063/5.0016505},
    url = {https://doi.org/10.1063/5.0016505},
    
}

@article{PhysRevLett.120.024102,
  title = {Model-Free Prediction of Large Spatiotemporally Chaotic Systems from Data: A Reservoir Computing Approach},
  author = {Pathak, Jaideep and Hunt, Brian and Girvan, Michelle and Lu, Zhixin and Ott, Edward},
  journal = {Phys. Rev. Lett.},
  volume = {120},
  issue = {2},
  pages = {024102},
  numpages = {5},
  year = {2018},
  month = {Jan},
  publisher = {American Physical Society},
  doi = {10.1103/PhysRevLett.120.024102},
  url = {https://link.aps.org/doi/10.1103/PhysRevLett.120.024102}
}

@article{10.1063/1.5120538,
    author = {Gao, Zhong-Ke and Li, Yan-Li and Yang, Yu-Xuan and Ma, Chao},
    title = "{A recurrence network-based convolutional neural network for fatigue driving detection from EEG}",
    journal = {Chaos: An Interdisciplinary Journal of Nonlinear Science},
    volume = {29},
    number = {11},
    pages = {113126},
    year = {2019},
    month = {11},
    
    issn = {1054-1500},
    doi = {10.1063/1.5120538},
    url = {https://doi.org/10.1063/1.5120538},
  
}

@article{LUKOSEVICIUS2009127,
title = {Reservoir computing approaches to recurrent neural network training},
journal = {Computer Science Review},
volume = {3},
number = {3},
pages = {127-149},
year = {2009},
issn = {1574-0137},
doi = {https://doi.org/10.1016/j.cosrev.2009.03.005},
url = {https://www.sciencedirect.com/science/article/pii/S1574013709000173},
author = {Mantas Lukoševičius and Herbert Jaeger}

}

@article{
doi:10.1126/science.1091277,
author = {Herbert Jaeger  and Harald Haas },
title = {Harnessing Nonlinearity: Predicting Chaotic Systems and Saving Energy in Wireless Communication},
journal = {Science},
volume = {304},
number = {5667},
pages = {78-80},
year = {2004},
doi = {10.1126/science.1091277},
URL = {https://www.science.org/doi/abs/10.1126/science.1091277}
}

@article{10.1063/1.4979665,
    author = {Lu, Zhixin and Pathak, Jaideep and Hunt, Brian and Girvan, Michelle and Brockett, Roger and Ott, Edward},
    title = "{Reservoir observers: Model-free inference of unmeasured variables in chaotic systems}",
    journal = {Chaos: An Interdisciplinary Journal of Nonlinear Science},
    volume = {27},
    number = {4},
    pages = {041102},
    year = {2017},
    month = {04},
    
    issn = {1054-1500},
    doi = {10.1063/1.4979665},
    url = {https://doi.org/10.1063/1.4979665},

}

@article{PhysRevE.98.052209,
  title = {Using reservoir computers to distinguish chaotic signals},
  author = {Carroll, T. L.},
  journal = {Phys. Rev. E},
  volume = {98},
  issue = {5},
  pages = {052209},
  numpages = {7},
  year = {2018},
  month = {Nov},
  publisher = {American Physical Society},
  doi = {10.1103/PhysRevE.98.052209},
  url = {https://link.aps.org/doi/10.1103/PhysRevE.98.052209}
}

@article{PhysRevE.104.014205,
  title = {Predicting amplitude death with machine learning},
  author = {Xiao, Rui and Kong, Ling-Wei and Sun, Zhong-Kui and Lai, Ying-Cheng},
  journal = {Phys. Rev. E},
  volume = {104},
  issue = {1},
  pages = {014205},
  numpages = {10},
  year = {2021},
  month = {Jul},
  publisher = {American Physical Society},
  doi = {10.1103/PhysRevE.104.014205},
  url = {https://link.aps.org/doi/10.1103/PhysRevE.104.014205}
}

@article{PhysRevResearch.3.013090,
  title = {Machine learning prediction of critical transition and system collapse},
  author = {Kong, Ling-Wei and Fan, Hua-Wei and Grebogi, Celso and Lai, Ying-Cheng},
  journal = {Phys. Rev. Res.},
  volume = {3},
  issue = {1},
  pages = {013090},
  numpages = {14},
  year = {2021},
  month = {Jan},
  publisher = {American Physical Society},
  doi = {10.1103/PhysRevResearch.3.013090},
  url = {https://link.aps.org/doi/10.1103/PhysRevResearch.3.013090}
}

@article{SAXENA2012205,
title = {Amplitude death: The emergence of stationarity in coupled nonlinear systems},
journal = {Physics Reports},
volume = {521},
number = {5},
pages = {205-228},
year = {2012},
note = {Amplitude Death: The Emergence of Stationarity in Coupled Nonlinear Systems},
issn = {0370-1573},
doi = {https://doi.org/10.1016/j.physrep.2012.09.003},

author = {Garima Saxena and Awadhesh Prasad and Ram Ramaswamy}
}

@article{KOSESKA2013173,
title = {Oscillation quenching mechanisms: Amplitude vs. oscillation death},
journal = {Physics Reports},
volume = {531},
number = {4},
pages = {173-199},
year = {2013},
note = {Oscillation quenching mechanisms: Amplitude vs. oscillation death},
issn = {0370-1573},
doi = {https://doi.org/10.1016/j.physrep.2013.06.001},
url = {https://www.sciencedirect.com/science/article/pii/S0370157313002652},
author = {Aneta Koseska and Evgeny Volkov and Jürgen Kurths}
}

@article{GREBOGI1983181,
title = {Crises, sudden changes in chaotic attractors, and transient chaos},
journal = {Physica D: Nonlinear Phenomena},
volume = {7},
number = {1},
pages = {181-200},
year = {1983},
issn = {0167-2789},
doi = {https://doi.org/10.1016/0167-2789(83)90126-4},
url = {https://www.sciencedirect.com/science/article/pii/0167278983901264}

}

@article{erdos1959random,
  title={On Random Graphs I},
  author={Erd{\H{o}}s, Paul and R{\'e}nyi, Alfréd},
  journal={Publicationes Mathematicae Debrecen},
  volume={6},
  pages={290--297},
  year={1959}
}

@article{ARONSON1990403,
title = {Amplitude response of coupled oscillators},
journal = {Physica D: Nonlinear Phenomena},
volume = {41},
number = {3},
pages = {403-449},
year = {1990},
issn = {0167-2789},
doi = {https://doi.org/10.1016/0167-2789(90)90007-C},
url = {https://www.sciencedirect.com/science/article/pii/016727899090007C},
author = {D.G. Aronson and G.B. Ermentrout and N. Kopell}

}

@article{10.1115/1.1579454,
    author = {Ermentrout,  B , Author and Mahajan,  A , Reviewer},
    title = "{Simulating, Analyzing, and Animating Dynamical Systems: A Guide to XPPAUT for Researchers and Students}",
    journal = {Applied Mechanics Reviews},
    volume = {56},
    number = {4},
    pages = {B53-B53},
    year = {2003},
    month = {07},
    issn = {0003-6900},
    doi = {10.1115/1.1579454},
    url = {https://doi.org/10.1115/1.1579454},
    
}

@ARTICLE{Kuznetsov:2008,
AUTHOR = {Kuznetsov, Y. A. and Sacker, R. J.},
TITLE   = {{N}eimark-{S}acker bifurcation},
YEAR    = {2008},
JOURNAL = {Scholarpedia},
VOLUME  = {3},
NUMBER  = {5},
PAGES   = {1845},
DOI     = {10.4249/scholarpedia.1845},
NOTE    = {revision \#91556}
}

@Inbook{Kuznetsov2023,
author="Kuznetsov, Yuri A.",
title="Bifurcations of Equilibria and Periodic Orbits in n-Dimensional Dynamical Systems",
bookTitle="Elements of Applied Bifurcation Theory",
year="2023",
publisher="Springer International Publishing",
address="Cham",
pages="175--228",
isbn="978-3-031-22007-4",
doi="10.1007/978-3-031-22007-4_5",
url="https://doi.org/10.1007/978-3-031-22007-4_5"
}

@article{LUO2006154,
title = {Double Neimark–Sacker bifurcation and torus bifurcation of a class of vibratory systems with symmetrical rigid stops},
journal = {Journal of Sound and Vibration},
volume = {298},
number = {1},
pages = {154-179},
year = {2006},
issn = {0022-460X},
doi = {https://doi.org/10.1016/j.jsv.2006.04.025},
url = {https://www.sciencedirect.com/science/article/pii/S0022460X06003907},
author = {G.W. Luo and Y.D. Chu and Y.L. Zhang and J.G. Zhang}
}

@article{article,
author = {Kuznetsov, Yu and Meijer, Hil},
year = {2006},
month = {10},
pages = {1009-1035},
title = {Remarks on interacting Neimark–Sacker bifurcations},
volume = {12},
journal = {Journal of Difference Equations and Applications},
doi = {10.1080/10236190600986842}
}

@article{PhysRevE.69.016207,
  title = {Statistics of finite-time Lyapunov exponents in the Ulam map},
  author = {Anteneodo, Celia},
  journal = {Phys. Rev. E},
  volume = {69},
  issue = {1},
  pages = {016207},
  numpages = {6},
  year = {2004},
  month = {Jan},
  publisher = {American Physical Society},
  doi = {10.1103/PhysRevE.69.016207},
  url = {https://link.aps.org/doi/10.1103/PhysRevE.69.016207}
}

@article{PhysRevE.60.2761,
  title = {Characteristic distributions of finite-time Lyapunov exponents},
  author = {Prasad, Awadhesh and Ramaswamy, Ramakrishna},
  journal = {Phys. Rev. E},
  volume = {60},
  issue = {3},
  pages = {2761--2766},
  numpages = {0},
  year = {1999},
  month = {Sep},
  publisher = {American Physical Society},
  doi = {10.1103/PhysRevE.60.2761},
  url = {https://link.aps.org/doi/10.1103/PhysRevE.60.2761}
}

@article{PhysRevLett.48.1507,
  title = {Chaotic Attractors in Crisis},
  author = {Grebogi, Celso and Ott, Edward and Yorke, James A.},
  journal = {Phys. Rev. Lett.},
  volume = {48},
  issue = {22},
  pages = {1507--1510},
  numpages = {0},
  year = {1982},
  month = {May},
  publisher = {American Physical Society},
  doi = {10.1103/PhysRevLett.48.1507},
  url = {https://link.aps.org/doi/10.1103/PhysRevLett.48.1507}
}

@article{RevModPhys.53.643,
  title = {Roads to turbulence in dissipative dynamical systems},
  author = {Eckmann, J. -P.},
  journal = {Rev. Mod. Phys.},
  volume = {53},
  issue = {4},
  pages = {643--654},
  numpages = {0},
  year = {1981},
  month = {Oct},
  publisher = {American Physical Society},
  doi = {10.1103/RevModPhys.53.643},
  url = {https://link.aps.org/doi/10.1103/RevModPhys.53.643}
}

@article{Kong_2021,
doi = {10.1088/2632-072X/ac0b00},
url = {https://doi.org/10.1088/2632-072X/ac0b00},
year = {2021},
month = {jul},
publisher = {IOP Publishing},
volume = {2},
number = {3},
pages = {035014},
author = {Kong, Ling-Wei and Fan, Huawei and Grebogi, Celso and Lai, Ying-Cheng},
title = {Emergence of transient chaos and intermittency in machine learning},
journal = {Journal of Physics: Complexity}
}

@book{Ott2002_9,
  author    = {Ott, Edward},
  title     = {Chaotic Transitions},
  booktitle = {Chaos in Dynamical Systems},
  publisher = {Cambridge University Press},
  address   = {Cambridge},
  year      = {2002},
  pages     = {345--377}
}

@article{PhysRevE.110.034211,
  title = {Dynamical analysis of a parameter-aware reservoir computer},
  author = {Sisodia, Dishant and Jalan, Sarika},
  journal = {Phys. Rev. E},
  volume = {110},
  issue = {3},
  pages = {034211},
  numpages = {7},
  year = {2024},
  month = {Sep},
  publisher = {American Physical Society},
  doi = {10.1103/PhysRevE.110.034211},
  url = {https://link.aps.org/doi/10.1103/PhysRevE.110.034211}
}

@article{8b7z-fnjd,
  title = {Dynamics of reservoir computing for crises prediction},
  author = {Sisodia, Dishant and Jalan, Sarika},
  journal = {Phys. Rev. E},
  pages = {--},
  year = {2026},
  month = {Apr},
  publisher = {American Physical Society},
  doi = {10.1103/8b7z-fnjd},
  url = {https://link.aps.org/doi/10.1103/8b7z-fnjd}
}

@article{Laiftle,
    author = {LAI, YING-CHENG},
    title = {EXTREME FLUCTUATIONS OF FINITE-TIME LYAPUNOV EXPONENTS IN CHAOTIC SYSTEMS},
    journal = {International Journal of Modern Physics B},
    volume = {21},
    number = {23n24},
    pages = {3941-3946},
    year = {2007},
    doi = {10.1142/S0217979207044998},
    
    URL = {https://doi.org/10.1142/S0217979207044998},
    eprint = {https://doi.org/10.1142/S0217979207044998}}

@incollection{Tsironis2025,
  author    = {Tsironis, G.},
  title     = {Complex Systems and Machine Learning},
  booktitle = {Artificial Intelligence and Complex Dynamical Systems},
  series    = {Understanding Complex Systems},
  publisher = {Springer},
  address   = {Cham},
  year      = {2025},
  doi       = {10.1007/978-3-031-81946-9_1}
}

@article{JaegerHaas2004,
  author  = {Jaeger, H. and Haas, H.},
  title   = {Harnessing Nonlinearity: Predicting Chaotic Systems and Saving Energy in Wireless Communication},
  journal = {Science},
  volume  = {304},
  number  = {5667},
  pages   = {78--80},
  year    = {2004},
  doi     = {10.1126/science.1091277}
}

@article{shrimali,
  title = {Predicting synchronization and oscillation death with parallel reservoir computing},
  author = {Chauhan, Swati and Verma, Umesh Kumar and Mandal, Swarnendu and Shrimali, Manish Dev},
  journal = {Phys. Rev. E},
  volume = {112},
  issue = {5},
  pages = {054220},
  numpages = {8},
  year = {2025},
  month = {Nov},
  publisher = {American Physical Society},
  doi = {10.1103/9n6x-qkpv},
  url = {https://link.aps.org/doi/10.1103/9n6x-qkpv}
}

@article{mandal,
    author = {Chauhan, Swati and Mandal, Swarnendu and Yadav, Vijay and Jaiswal, Prabhat K. and Priya, Madhu and Shrimali, Manish Dev},
    title = {Machine learning based prediction of phase ordering dynamics},
    journal = {Chaos: An Interdisciplinary Journal of Nonlinear Science},
    volume = {33},
    number = {6},
    pages = {061103},
    year = {2023},
    month = {06},
    
    issn = {1054-1500},
    doi = {10.1063/5.0156611},
    url = {https://doi.org/10.1063/5.0156611}
 }

@article{Brunton2016,
  author  = {Brunton, Steven L. and Proctor, Joshua L. and Kutz, J. Nathan},
  title   = {Discovering governing equations from data by sparse identification of nonlinear dynamical systems},
  journal = {Proceedings of the National Academy of Sciences of the United States of America},
  volume  = {113},
  number  = {15},
  pages   = {3932--3937},
  year    = {2016},
  doi     = {10.1073/pnas.1517384113}
}

@article{10.1063/5.0283386,
    author = {Inoue, Katsuma and Kubota, Tomoyuki and Hoan Tran, Quoc and Akashi, Nozomi and Terajima, Ryo and Kabayama, Tempei and Guan, JingChuan and Nakajima, Kohei},
    title = {Reservoir computing bootcamp—From Python/NumPy tutorial for the complete beginners to cutting-edge research topics of reservoir computing},
    journal = {Chaos: An Interdisciplinary Journal of Nonlinear Science},
    volume = {36},
    number = {2},
    pages = {023109},
    year = {2026},
    month = {02},
    issn = {1054-1500},
    doi = {10.1063/5.0283386},
    url = {https://doi.org/10.1063/5.0283386}
}

@article{10.1063/5.0252908,
    author = {Hramov, Alexander E. and Kulagin, Nikita and Pisarchik, Alexander N. and Andreev, Andrey V.},
    title = {Strong and weak prediction of stochastic dynamics using reservoir computing},
    journal = {Chaos: An Interdisciplinary Journal of Nonlinear Science},
    volume = {35},
    number = {3},
    pages = {033140},
    year = {2025},
    month = {03},
    doi = {10.1063/5.0252908},
    url = {https://doi.org/10.1063/5.0252908}
}

@article{Bury2023,
  author  = {Bury, T. M. and Dylewsky, D. and Bauch, C. T. and others},
  title   = {Predicting discrete-time bifurcations with deep learning},
  journal = {Nature Communications},
  volume  = {14},
  pages   = {6331},
  year    = {2023},
  doi     = {10.1038/s41467-023-42020-z}
}

@article{Choudhary2020,
  author  = {Choudhary, A. and Lindner, J. F. and Holliday, E. G. and Miller, S. T. and Sinha, S. and Ditto, W. L.},
  title   = {Physics-enhanced neural networks learn order and chaos},
  journal = {Physical Review E},
  volume  = {101},
  number  = {6},
  pages   = {062207},
  year    = {2020},
  doi     = {10.1103/PhysRevE.101.062207}
}

@article{Bollt2021,
  author  = {Bollt, E.},
  title   = {On explaining the surprising success of reservoir computing forecaster of chaos? The universal machine learning dynamical system with contrast to VAR and DMD},
  journal = {Chaos},
  volume  = {31},
  number  = {1},
  pages   = {013108},
  year    = {2021},
  doi     = {10.1063/5.0024890}
}

@article{PhysRevResearch.6.043194,
  author  = {Panahi, S. and Kong, L.-W. and Moradi, M. and Zhai, Z.-M. and Glaz, B. and Haile, M. and Lai, Y.-C.},
  title   = {Machine learning prediction of tipping in complex dynamical systems},
  journal = {Phys. Rev. Res.},
  volume  = {6},
  pages   = {043194},
  year    = {2024},
  doi     = {10.1103/PhysRevResearch.6.043194}
}

@article{8016349,
  author  = {Katz, G. E. and Reggia, J. A.},
  title   = {Using Directional Fibers to Locate Fixed Points of Recurrent Neural Networks},
  journal = {IEEE Transactions on Neural Networks and Learning Systems},
  volume  = {29},
  number  = {8},
  pages   = {3636--3646},
  year    = {2018},
  doi     = {10.1109/TNNLS.2017.2733544}
}

@misc{directional-fibers,
  author       = {Katz, G.},
  title        = {Directional Fibers},
  year         = {2023},
  howpublished = {Computer software, GitHub repository},
  url          = {https://github.com/garrettkatz/directional-fibers}
}

@article{ananthkrishnan2001crises,
  author  = {Ananthkrishnan, N. and Sahai, T.},
  title   = {Crises},
  journal = {Resonance},
  volume  = {6},
  pages   = {19--33},
  year    = {2001},
  doi     = {10.1007/BF02837669}
}

@inbook{Ott2002_8,
  author    = {Ott, E.},
  title     = {Chaotic transitions},
  booktitle = {Chaos in Dynamical Systems},
  publisher = {Cambridge University Press},
  address   = {Cambridge},
  year      = {2002},
  pages     = {304--344}
}

@article{Inubushi2017,
  author    = {Inubushi, Masanobu and Yoshimura, Kazuyuki},
  title     = {Reservoir Computing Beyond Memory-Nonlinearity Trade-off},
  journal   = {Scientific Reports},
  volume    = {7},
  number    = {1},
  pages     = {10199},
  year      = {2017},
  month     = {Aug},
  doi       = {10.1038/s41598-017-10257-6},
  url       = {https://doi.org/10.1038/s41598-017-10257-6}
}

@article{PhysRevE.87.042809,
  title = {Mean-field theory of echo state networks},
  author = {Massar, Marc and Massar, Serge},
  journal = {Phys. Rev. E},
  volume = {87},
  issue = {4},
  pages = {042809},
  numpages = {7},
  year = {2013},
  month = {Apr},
  publisher = {American Physical Society},
  doi = {10.1103/PhysRevE.87.042809},
  url = {https://link.aps.org/doi/10.1103/PhysRevE.87.042809}
}

@article{Carroll2020,
  author    = {Carroll, Thomas L.},
  title     = {Do reservoir computers work best at the edge of chaos?},
  journal   = {Chaos: An Interdisciplinary Journal of Nonlinear Science},
  volume    = {30},
  number    = {12},
  pages     = {121109},
  year      = {2020},
  month     = {Dec},
  doi       = {10.1063/5.0038163},
  url       = {https://doi.org/10.1063/5.0038163}
}

@article{Mizzi2025,
  author    = {Mizzi, Antony and Small, Michael and Walker, David M.},
  title     = {Reservoir computing with the minimum description length principle},
  journal   = {Chaos: An Interdisciplinary Journal of Nonlinear Science},
  volume    = {35},
  number    = {4},
  pages     = {043132},
  year      = {2025},
  month     = {Apr},
  doi       = {10.1063/5.0252938},
  url       = {https://doi.org/10.1063/5.0252938}
}

@article{10.1016/j.neunet.2007.04.001,
author = {Jaeger, Herbert and Maass, Wolfgang and Principe, Jose},
title = {Editorial: Special issue on echo state networks and liquid state machines},
year = {2007},
issue_date = {April, 2007},
publisher = {Elsevier Science Ltd.},
address = {GBR},
volume = {20},
number = {3},
issn = {0893-6080},
url = {https://doi.org/10.1016/j.neunet.2007.04.001},
doi = {10.1016/j.neunet.2007.04.001},
journal = {Neural Netw.},
month = apr,
pages = {287–289},
numpages = {3}
}

@article{PhysRevFluids.5.113506,
  title = {Reservoir computing model of two-dimensional turbulent convection},
  author = {Pandey, Sandeep and Schumacher, J\"org},
  journal = {Phys. Rev. Fluids},
  volume = {5},
  issue = {11},
  pages = {113506},
  numpages = {18},
  year = {2020},
  month = {Nov},
  publisher = {American Physical Society},
  doi = {10.1103/PhysRevFluids.5.113506},
  url = {https://link.aps.org/doi/10.1103/PhysRevFluids.5.113506}
}

@article{https://doi.org/10.1029/2020MS002290,
author = {Nadiga, Balasubramanya T.},
title = {Reservoir Computing as a Tool for Climate Predictability Studies},
journal = {Journal of Advances in Modeling Earth Systems},
volume = {13},
number = {4},
pages = {e2020MS002290},
doi = {https://doi.org/10.1029/2020MS002290},

year = {2021}
}

@article{SANTOS2025134698,
title = {Reservoir computing and non-linear dynamics for time series analysis: An application in the financial market},
journal = {Physica D: Nonlinear Phenomena},
volume = {476},
pages = {134698},
year = {2025},
issn = {0167-2789},
doi = {https://doi.org/10.1016/j.physd.2025.134698},
url = {https://www.sciencedirect.com/science/article/pii/S0167278925001757},
author = {Francisco Alves dos Santos and Reneé Rodrigues Lima and Jerson Leite Alves and Davi Wanderley Misturini and Joao B. Florindo}
}

@article{srinivasan2025boosting,
  author    = {Srinivasan, K. and Plenz, D. and Girvan, M.},
  title     = {Boosting reservoir computing with brain-inspired adaptive control of E-I balance},
  journal   = {Nature Communications},
  year      = {2025},
  volume    = {16},
  number    = {1},
  pages     = {10212},
  doi       = {10.1038/s41467-025-64978-8},
  url       = {https://doi.org/10.1038/s41467-025-64978-8},
  month     = {Nov}
}

@article{li2025reservoir,
  author    = {Li, Gen and Huang, Liang and Lei, Youming},
  title     = {Reservoir computing meeting Kolmogorov–Arnold networks: Prediction of high-dimensional chaotic systems},
  journal   = {Chaos: An Interdisciplinary Journal of Nonlinear Science},
  year      = {2025},
  volume    = {35},
  number    = {10},
  pages     = {103120},
  doi       = {10.1063/5.0289577},
  url       = {https://doi.org/10.1063/5.0289577},
  month     = {Oct}
}

@article{alam2026attractor,
  author    = {Alam, Mohammad Shah Alam and Ott, William and Timofeyev, Ilya},
  title     = {Attractor learning for spatiotemporally chaotic dynamical systems using echo state networks with transfer learning},
  journal   = {Chaos: An Interdisciplinary Journal of Nonlinear Science},
  year      = {2026},
  volume    = {36},
  number    = {2},
  pages     = {023105},
  doi       = {10.1063/5.0283121},
  url       = {https://doi.org/10.1063/5.0283121},
  month     = {Feb}
}

@article{ashwin2012tipping,
  author    = {Ashwin, Peter and Wieczorek, Sebastian and Vitolo, Renato and Cox, Peter},
  title     = {Tipping points in open systems: bifurcation, noise-induced and rate-dependent examples in the climate system},
  journal   = {Philosophical Transactions of the Royal Society A: Mathematical, Physical and Engineering Sciences},
  year      = {2012},
  volume    = {370},
  number    = {1962},
  pages     = {1166--1184},
  doi       = {10.1098/rsta.2011.0306},
  url       = {https://doi.org/10.1098/rsta.2011.0306},
  month     = {Mar}
}

@article{kong2023reservoir,
  author    = {Kong, Ling-Wei and Weng, Yang and Glaz, Bryan and Haile, Mulugeta and Lai, Ying-Cheng},
  title     = {Reservoir computing as digital twins for nonlinear dynamical systems},
  journal   = {Chaos: An Interdisciplinary Journal of Nonlinear Science},
  year      = {2023},
  volume    = {33},
  number    = {3},
  pages     = {033111},
  doi       = {10.1063/5.0138661},
  url       = {https://doi.org/10.1063/5.0138661},
  month     = {Mar}
}

@article{panahi2024adaptable,
  author    = {Panahi, Shirin and Lai, Ying-Cheng},
  title     = {Adaptable reservoir computing: A paradigm for model-free data-driven prediction of critical transitions in nonlinear dynamical systems},
  journal   = {Chaos: An Interdisciplinary Journal of Nonlinear Science},
  year      = {2024},
  volume    = {34},
  number    = {5},
  pages     = {051501},
  doi       = {10.1063/5.0200898},
  url       = {https://doi.org/10.1063/5.0200898},
  month     = {May}
}

@article{SHAHI2022100300,
title = {Prediction of chaotic time series using recurrent neural networks and reservoir computing techniques: A comparative study},
journal = {Machine Learning with Applications},
volume = {8},
pages = {100300},
year = {2022},
issn = {2666-8270},
doi = {https://doi.org/10.1016/j.mlwa.2022.100300},
url = {https://www.sciencedirect.com/science/article/pii/S2666827022000275},
author = {Shahrokh Shahi and Flavio H. Fenton and Elizabeth M. Cherry}


}

@InProceedings{pmlr-v199-cisneros22a,
  title = 	 {Benchmarking Learning Efficiency in Deep Reservoir Computing},
  author =       {Cisneros, Hugo and Mikolov, Tomas and Sivic, Josef},
  booktitle = 	 {Proceedings of The 1st Conference on Lifelong Learning Agents},
  pages = 	 {532--547},
  year = 	 {2022},
  editor = 	 {Chandar, Sarath and Pascanu, Razvan and Precup, Doina},
  volume = 	 {199},
  series = 	 {Proceedings of Machine Learning Research},
  month = 	 {22--24 Aug},
  publisher =    {PMLR},
  
  url = 	 {https://proceedings.mlr.press/v199/cisneros22a.html}
  
}

@Article{npg-27-373-2020,
AUTHOR = {Chattopadhyay, A. and Hassanzadeh, P. and Subramanian, D.},
TITLE = {Data-driven predictions of a multiscale Lorenz 96 chaotic system using machine-learning methods: reservoir computing, artificial neural network, and long short-term memory network},
JOURNAL = {Nonlinear Processes in Geophysics},
VOLUME = {27},
YEAR = {2020},
NUMBER = {3},
PAGES = {373--389},
URL = {https://npg.copernicus.org/articles/27/373/2020/},
DOI = {10.5194/npg-27-373-2020}
}

@article{ROSENSTEIN199482,
title = {Reconstruction expansion as a geometry-based framework for choosing proper delay times},
journal = {Physica D: Nonlinear Phenomena},
volume = {73},
number = {1},
pages = {82-98},
year = {1994},
issn = {0167-2789},
doi = {https://doi.org/10.1016/0167-2789(94)90226-7},
url = {https://www.sciencedirect.com/science/article/pii/0167278994902267},
author = {Michael T. Rosenstein and James J. Collins and Carlo J. {De Luca}}
}

@article{tan2023selecting,
  author    = {Tan, Eugene and Algar, Shannon and Corr{\^e}a, D{\'e}bora and Small, Michael and Stemler, Thomas and Walker, David},
  title     = {Selecting embedding delays: An overview of embedding techniques and a new method using persistent homology},
  journal   = {Chaos: An Interdisciplinary Journal of Nonlinear Science},
  year      = {2023},
  volume    = {33},
  number    = {3},
  pages     = {032101},
  doi       = {10.1063/5.0137223},
  url       = {https://doi.org/10.1063/5.0137223},
  month     = {Mar}
}

@article{whitney1936differentiable,
  author    = {Whitney, Hassler},
  title     = {Differentiable manifolds},
  journal   = {Annals of Mathematics},
  year      = {1936},
  volume    = {37},
  number    = {3},
  pages     = {645--680},
  doi       = {10.2307/1968482},
  url       = {https://doi.org/10.2307/1968482},
  publisher = {Princeton University}
}

@inproceedings{takens1981detecting,
  author    = {Takens, Floris},
  title     = {Detecting strange attractors in turbulence},
  booktitle = {Dynamical Systems and Turbulence, Warwick 1980},
  series    = {Lecture Notes in Mathematics},
  volume    = {898},
  pages     = {366--381},
  year      = {1981},
  publisher = {Springer},
  address   = {Berlin, Heidelberg},
  doi       = {10.1007/BFb0091924}
}

@article{PhysRevLett.45.712,
  title = {Geometry from a Time Series},
  author = {Packard, N. H. and Crutchfield, J. P. and Farmer, J. D. and Shaw, R. S.},
  journal = {Phys. Rev. Lett.},
  volume = {45},
  issue = {9},
  pages = {712--716},
  numpages = {0},
  year = {1980},
  month = {Sep},
  publisher = {American Physical Society},
  doi = {10.1103/PhysRevLett.45.712},
  url = {https://link.aps.org/doi/10.1103/PhysRevLett.45.712}
}

@article{sisodia2026finite,
  author    = {Sisodia, D. and Jalan, S.},
  title     = {Finite-time Lyapunov signatures of dynamical regimes in reservoir models},
  journal   = {The European Physical Journal B},
  year      = {2026},
  volume    = {99},
  number    = {6},
  pages     = {85},
  doi       = {10.1140/epjb/s10051-026-01204-4},
  url       = {https://doi.org/10.1140/epjb/s10051-026-01204-4},
  month     = {Jun}
}

@article{GRIGORYEVA2018495,
title = {Echo state networks are universal},
journal = {Neural Networks},
volume = {108},
pages = {495-508},
year = {2018},
issn = {0893-6080},
doi = {https://doi.org/10.1016/j.neunet.2018.08.025},
url = {https://www.sciencedirect.com/science/article/pii/S089360801830251X},
author = {Lyudmila Grigoryeva and Juan-Pablo Ortega}
}

@article{HART2020234,
title = {Embedding and approximation theorems for echo state networks},
journal = {Neural Networks},
volume = {128},
pages = {234-247},
year = {2020},
issn = {0893-6080},
doi = {https://doi.org/10.1016/j.neunet.2020.05.013},
url = {https://www.sciencedirect.com/science/article/pii/S0893608020301830},
author = {Allen Hart and James Hook and Jonathan Dawes}
}

@article{PhysRevResearch.5.L022041,
  title = {Embedding theory of reservoir computing and reducing reservoir network using time delays},
  author = {Duan, Xing-Yue and Ying, Xiong and Leng, Si-Yang and Kurths, J\"urgen and Lin, Wei and Ma, Huan-Fei},
  journal = {Phys. Rev. Res.},
  volume = {5},
  issue = {2},
  pages = {L022041},
  numpages = {6},
  year = {2023},
  month = {May},
  publisher = {American Physical Society},
  doi = {10.1103/PhysRevResearch.5.L022041},
  url = {https://link.aps.org/doi/10.1103/PhysRevResearch.5.L022041}
}

@article{PhysRevLett.76.1816,
  title = {Generalized Synchronization, Predictability, and Equivalence of Unidirectionally Coupled Dynamical Systems},
  author = {Kocarev, L. and Parlitz, U.},
  journal = {Phys. Rev. Lett.},
  volume = {76},
  issue = {11},
  pages = {1816--1819},
  numpages = {0},
  year = {1996},
  month = {Mar},
  publisher = {American Physical Society},
  doi = {10.1103/PhysRevLett.76.1816},
  url = {https://link.aps.org/doi/10.1103/PhysRevLett.76.1816}
}

@INPROCEEDINGS{6706899,
  author={Caluwaerts, Ken and Wyffels, Francis and Dieleman, Sander and Schrauwen, Benjamin},
  booktitle={The 2013 International Joint Conference on Neural Networks (IJCNN)}, 
  title={The spectral radius remains a valid indicator of the Echo state property for large reservoirs}, 
  year={2013},
  volume={},
  number={},
  pages={1-6},
  doi={10.1109/IJCNN.2013.6706899}}

@incollection{HIRSCH2013329,
title = {15 - Discrete Dynamical Systems},
booktitle = {Differential Equations, Dynamical Systems, and an Introduction to Chaos (Third Edition)},
publisher = {Academic Press},
edition = {Third Edition},
address = {Boston},
pages = {329-359},
year = {2013},
isbn = {978-0-12-382010-5},
doi = {https://doi.org/10.1016/B978-0-12-382010-5.00015-4},
url = {https://www.sciencedirect.com/science/article/pii/B9780123820105000154},
author = {Morris W. Hirsch and Stephen Smale and Robert L. Devaney}
}

@software{reservoir_computing_tutorial_code,
  
  title        = {Supplementary Code for: Reservoir as a Dynamical System: A Tutorial Review},
  doi = {10.5281/zenodo.21778344},
  url          = {https://github.com/complex-systems-lab/Tutorial_Review_Reservoir_Computing}
}

@article{PhysRevE.110.034204,
  title = {Reservoir computing with logistic map},
  author = {Arun, R. and Sathish Aravindh, M. and Venkatesan, A. and Lakshmanan, M.},
  journal = {Phys. Rev. E},
  volume = {110},
  issue = {3},
  pages = {034204},
  numpages = {12},
  year = {2024},
  month = {Sep},
  publisher = {American Physical Society},
  doi = {10.1103/PhysRevE.110.034204},
  url = {https://link.aps.org/doi/10.1103/PhysRevE.110.034204}
}

@article{f2f3-l3fc,
  title = {Spin-torque nano-oscillators for reservoir computing applied to time-series data},
  author = {Arun, R. and Sathish Aravindh, M. and Lakshmanan, M.},
  journal = {Phys. Rev. E},
  volume = {112},
  issue = {1},
  pages = {015307},
  numpages = {11},
  year = {2025},
  month = {Jul},
  publisher = {American Physical Society},
  doi = {10.1103/f2f3-l3fc},
  url = {https://link.aps.org/doi/10.1103/f2f3-l3fc}
}

@article{PhysRevE.105.054203,
  title = {Machine-learning potential of a single pendulum},
  author = {Mandal, Swarnendu and Sinha, Sudeshna and Shrimali, Manish Dev},
  journal = {Phys. Rev. E},
  volume = {105},
  issue = {5},
  pages = {054203},
  numpages = {9},
  year = {2022},
  month = {May},
  publisher = {American Physical Society},
  doi = {10.1103/PhysRevE.105.054203},
  url = {https://link.aps.org/doi/10.1103/PhysRevE.105.054203}
}

@article{10.1063/5.0291585,
    author = {Mandal, Swarnendu and Chauhan, Swati and Verma, Umesh Kumar and Shrimali, Manish Dev and Aihara, Kazuyuki},
    title = {Adaptive control of dynamical systems using reservoir computing},
    journal = {Chaos: An Interdisciplinary Journal of Nonlinear Science},
    volume = {35},
    number = {9},
    pages = {091102},
    year = {2025},
    month = {09},
    issn = {1054-1500},
    doi = {10.1063/5.0291585},
    url = {https://doi.org/10.1063/5.0291585}
}

@article{10.1063/5.0119963,
    author = {Roy, Mousumi and Mandal, Swarnendu and Hens, Chittaranjan and Prasad, Awadhesh and Kuznetsov, N. V. and Dev Shrimali, Manish},
    title = {Model-free prediction of multistability using echo state network},
    journal = {Chaos: An Interdisciplinary Journal of Nonlinear Science},
    volume = {32},
    number = {10},
    pages = {101104},
    year = {2022},
    month = {10},
}

@article{10.1063/5.0271356,
    author = {Chauhan, Swati and Mandal, Swarnendu and Dixit, Shiva and Dev Shrimali, Manish},
    title = {Predicting collective states of a star network using reservoir computing},
    journal = {Chaos: An Interdisciplinary Journal of Nonlinear Science},
    volume = {35},
    number = {12},
    pages = {123112},
    year = {2025},
    month = {12},
    issn = {1054-1500},
    doi = {10.1063/5.0271356},
    url = {https://doi.org/10.1063/5.0271356},
}

\end{document}